\documentclass[twocolumn]{aastex631}
\usepackage{amsmath}

\shorttitle{The population of NuSTAR BH XBs}
\shortauthors{Draghis et al.}

\begin{document}

\title{The population of NuSTAR Black Hole X-ray Binaries}
\correspondingauthor{Paul A. Draghis}
\author[0000-0002-2218-2306]{Paul A. Draghis}
\email{pdraghis@mit.edu}
\affiliation{MIT Kavli Institute for Astrophysics and Space Research, Massachusetts Institute of Technology, 70 Vassar St, Cambridge, MA 02139, USA}
\affiliation{Department of Astronomy, University of Michigan, 1085 South University Avenue, Ann Arbor, MI 48109, USA}

\author[0000-0003-2869-7682]{Jon M. Miller}
\affiliation{Department of Astronomy, University of Michigan, 1085 South University Avenue, Ann Arbor, MI 48109, USA}


\author[0000-0003-2663-1954]{Laura Brenneman}
\affiliation{Center for Astrophysics, Harvard \& Smithsonian, 60 Garden Street, Cambridge, MA 02138, USA}

\author[0000-0001-8470-749X]{Elisa Costantini}
\affiliation{SRON Netherlands Institute for Space Research, Niels Bohrweg 4, 2333 CA Leiden, The Netherlands}
\affiliation{Anton Pannekoek Astronomical Institute, University of Amsterdam, P.O. Box 94249, 1090 GE Amsterdam, The Netherlands}

\author[0009-0006-4968-7108]{Luigi C. Gallo}
\affiliation{Department of Astronomy \& Physics, Saint Mary’s University, 923 Robie Street, Halifax, Nova Scotia, B3H 3C3, Canada}


\author[0000-0003-1621-9392]{Mark Reynolds}
\affiliation{Department of Astronomy, Ohio State University,  140 W 18th Avenue, Columbus, OH 43210, USA}

\author[0000-0001-5506-9855]{John A. Tomsick}
\affiliation{Space Sciences Laboratory, 7 Gauss Way, University of California, Berkeley, CA 94720-7450, USA}

\author[0000-0002-0572-9613]{Abderahmen Zoghbi}
\affiliation{Department of Astronomy, University of Maryland, College Park, MD 20742, USA}
\affiliation{HEASARC, Code 6601, NASA/GSFC, Greenbelt, MD 20771, USA}
\affiliation{CRESST II, NASA Goddard Space Flight Center, Greenbelt, MD 20771, USA}


\begin{abstract}

The spin of a black hole (BH) encodes information about its formation and evolution history. Yet the understanding of the distribution of BH spins in X-ray binaries (XBs), of the models used to measure spin, and of their impact on systematic uncertainties remains incomplete. In this work, we expand on previous analyses of the entire NuSTAR archive of accreting BH XBs. Prior work compiled a sample of 245 spectral fits using the relativistic reflection method for NuSTAR observations of 36 BH systems. Here, we aim to probe two aspects: the connection between BH spin and binary system properties, and the relationships between parameters in the spectral fits. We identify moderate negative correlations between spin uncertainty and both BH mass and system inclination, and a moderate positive correlation with distance. We also point out tentative multidimensional degeneracies between inclination, disk density, Fe abundance, ionization, and the presence or absence of absorption features from ionized outflows linked to disk winds. Lastly, we provide a comprehensive view of the observed distribution of BH spins in XBs, in comparison to spins inferred from gravitational waves. We find that the distribution of BH spins in XBs can be described by a beta distribution with $\alpha=5.66$ and $\beta=1.09$. This data set is highly complex, and the analysis presented here does not fully explore all potential parameter correlations. We make the full data set available in Zenodo to the community to encourage further exploration.
\end{abstract}


\section{Introduction} \label{sec:intro}

The formation mechanisms and evolutionary paths of black holes (BHs) are unclear. This is despite the theoretical simplicity of BHs, which are defined solely by their mass and rotation, expressed as the dimensionless spin parameter $a$ ($-1<a=cJ/GM^2<1$)\footnote{Where c is speed of light, G is the gravitational constant, and J and M are the angular momentum and mass of the BH.}. Stellar-mass BHs originate from massive stars. As these stars are often part of binary or multiple systems, BH formation must be understood in relation to the host binary system. Recent numerical simulations suggest that isolated BHs have low natal spins (\citealt{2004ApJ...602..312G, 2022MNRAS.511..176A}). However, similar efforts for BHs forming in binary systems produce uncertain results, heavily influenced by assumptions regarding stellar evolution and binary interactions (see, e.g., \citealt{2022ApJ...925...69B}). Given that BHs only have two observable properties, understanding BH rotation is essential for any attempt to explain how BHs form and the impact they have on their surroundings on small and large scales.

Because they do not emit electromagnetic radiation, studying BHs requires probing the influence they have on their surrounding media. In the case of stellar-mass BHs, X-ray binary (XB) systems are ideal laboratories for studying the behavior of matter in regions in the proximity of the BH. In these systems, a stellar-mass BH is accreting matter from a stellar binary companion. The most common classification of XBs separates them into high-mass XBs (HMXBs) and low-mass XBs (LMXBs). In the former case, the stellar companion is a massive star (with a typical mass of a few solar masses) that feeds the accretion disk around the BH through stellar winds. In the latter, the accretion disk is formed by material leaving the typically subsolar-mass stellar companion through Roche lobe overflow. These systems are most frequently discovered when entering an ``outburst" period, when their X-ray emission increases by many orders of magnitude, hence the name: X-ray binaries. Still, the mechanism that causes an XB to enter an outburst remains elusive. Spectral studies of accreting BHs have evolved significantly over the era of X-ray astronomy through the multitude of observatories that have operated over the past few decades. Consequently, the work presented here represents a significant effort aimed at expanding our understanding of accreting BH XBs through X-ray spectral studies.

\subsection{X-ray Spin Measurement Techniques}

X-ray spectral characterization has emerged as the standard technique in the field to probe BH spin. All X-ray measurements of BH spin assume an optically thick, geometrically thin accretion disk that extends close to the innermost stable circular orbit (ISCO). The size of the ISCO is determined by the spin of the BH: as the spin increases between its theoretical limits of -1 and 1, the ISCO radius decreases from 9 $r_g$ to 1 $r_g$. In particular, two methods are most commonly used: ``disk continuum fitting" (see, e.g., \citealt{2009ApJ...701.1076G, 2012ApJ...744..107N, 2014ApJ...793L..29S}) and ``relativistic reflection" (see, e.g. \citealt{2002ApJ...578..348M, 2006ApJ...652.1028B, 2009Natur.459..540F}).

The continuum fitting method aims to characterize the shape of the emission from the thermal accretion disk by considering the effects that the BH rotation has on the spectral profile. In order to break parameter degeneracies in the models, this method relies on the existence of prior, independent knowledge regarding the mass of the BH, the distance to the system, and the viewing inclination of the inner regions of the accretion disk. BH masses and inclinations are often inferred based on optical spectroscopic measurements (see, e.g., \citealt{2014ApJ...784....2M, 2017ApJ...846..132H, 2021ApJ...920..120Y, 2025AA...694A.119Y}). However, it is often challenging to observe the stellar companions owing to the systems generally being located at large distances, which both decrease the flux and lead to significant absorption due to the interstellar medium. In terms of distances, parallax measurements represent the most pragmatic tool, as most other methods are dependent on model assumptions (\citealt{2008JPhCS.131a2057M}). Radio parallax measurements have produced reliable constraints for a number of BH XBs up to $\sim 10\;\rm kpc$ (see, e.g., \citealt{2009ApJ...706L.230M, 2014ApJ...796....2R, 2020MNRAS.493L..81A, 2021Sci...371.1046M}), but still the precision of the measurements decreases with increasing distance. Furthermore, using the inclination determined from the properties of the binary system in continuum fitting measurements assumes that the angular momentum of the BH is aligned with the angular momentum of the binary. However, if natal kicks are present during the supernova events that produce the BHs in the systems, the BH can be misaligned when compared to the binary. Due to the Bardeen-Petterson effect (\citealt{1975ApJ...195L..65B}), the angular momentum of the inner regions of the accretion disk is expected to align with the angular momentum of the BH. If the BH is misaligned, this would lead to different viewing inclinations for the inner accretion disk, responsible for the emission characterized through continuum fitting, and for the binary system. Lastly, continuum fitting is strongly based on assumptions regarding the spectral hardening factor ($f_{\rm col}$; \citealt{1995ApJ...445..780S}). A canonical value of 1.7 is often used, but studies such as \cite{2013MNRAS.431.3510S} described how allowing the hardening factor to vary can fully describe the variability seen in the disk properties in such systems. Continuum fitting, while indeed a powerful technique, suffers from the limitations set by the requirement of independent information about the system and by assumptions regarding physical interpretations. For these reasons, only $\sim10$ systems have continuum fitting spin measurements with uncertainty $\leq10\%$.

In contrast, relativistic reflection is a technique that models distortions of spectral features in a relative manner, rather than based on absolute fluxes. Therefore, this method does not require prior information regarding the observed systems, such as the BH mass or distance, making it a tool that is applicable to BHs across the entire mass range and ideally suited for studies of systems where information outside of X-ray observations is limited or nonexistent. This enables reflection studies of stellar-mass BHs (\citealt{2006ApJ...652.1028B, 2009ApJ...697..900M}) and of supermassive BHs in active galactic nuclei (AGN; \citealt{2015MNRAS.446..633G, 2015ApJ...806..149K}).  However, reflection studies are likely to be limited by assumptions that are built into the models that stem from an incomplete characterization of the physical mechanisms at play in the systems. This includes an incomplete understanding of the geometry of the accretion disk and of the corona, including the vertical extent of accretion disk, changes in the inner and outer disk radius, the presence of disk warps, and the time evolution of the coronal location, size, and temperature. Furthermore, it is unclear how the assumption regarding the angular distribution of the incident emission in the models, the assumption of constant ionization and density of the atmosphere of the accretion disk, and the proper characterization of the elemental abundances in the accretion disk impact the spin measurement. Lastly, the ability to disentangle the reflected emission from the underlying continuum and the impact of narrow emission and absorption features, together with complex partial obscuration and distant reflection, is crucial for properly characterizing the complex shape of the reflection spectrum, and therefore the quality of the spin measurement. For these reasons, understanding the behavior of the reflection models is crucial for characterizing the robustness of the spin measurements using the relativistic reflection method. This work aims to probe the impact of some of these effects, especially when applied to current-generation data.

The 2012 launch of NuSTAR (\citealt{2013ApJ...770..103H}) revolutionized BH spin measurements by providing unrivaled sensitivity and a wide bandpass that covers the main features of relativistic reflection: the Fe K complex around 6.4 keV and the Compton hump around 20 keV. The current explanation for spectra of XB systems involves three main components: emission from a hot, thermal, blackbody-like accretion disk that is fueled by material flowing from a companion star, a power-law component with a high-energy cutoff describing the Comptonized radiation from a hot compact ``corona," and radiation ``reflected" by the atmosphere of the innermost regions of the accretion disk, carrying information about the BH. The latter, responsible for BH spin measurements, consists of spectral features that are relativistically broadened by the BH's extreme gravity. As matter orbits the BH closer to its event horizon, gravitational effects will become stronger, imprinting more significant features on the radiation, such as Doppler shifts, gravitational redshift, Doppler beaming, and gravitational lensing. By measuring the distortion of these spectral features, we can quantify the gravitational effects of the BH and infer the proximity of the emitting regions to the BH. By further equating this with the size of the ISCO of the BH, we obtain a direct probe of the BH spin. 
\newline
\subsection{Recent Developments in Black Hole Spin Measurements}

As highlighted in \cite{Draghis24}, the extent of systematic uncertainties of X-ray measurements of BH spins represents a major discussion point in the field. Key sources of systematic uncertainty include (1) discrepancies among model families or variations within a model family, due to assumptions of the elemental abundances or photoelectric cross sections; (2) the validity of the assumptions about the physical processes in the observed systems and the precision of models in representing source behavior; (3) unmodeled small-scale source variability; and (4) incomplete understanding of models and potential parameter correlations leading to unexplored parameter spaces. Moreover, in characterizing the BH spin population in XBs, it is essential to note that only a relatively small number of these systems have been detected in our Galaxy, factoring in potential selection effects and observational biases.

In recent years, significant efforts have been made to improve the robustness of XB BH spin measurements, both from the data and model perspective. \cite{2023ApJ...954...62D} emphasize the importance of analyzing multiple observations of the same source in trying to reliably quantify the magnitude of the systematic uncertainties of the measurements. Works such as \cite{2024A&A...684A..95M} and \cite{2024ApJ...960....3S} include X-ray polarization information from the IXPE telescope (\citealt{2022JATIS...8b6002W}) in the spin measurements of BHs in XBs. \cite{2024FrASS..1092682C} demonstrate the improvement in spin measurements that would be brought forth by an instrument with a large effective area, a broad bandpass that covers both the soft and hard X-ray regimes, and improved energy resolution. Simultaneously, significant progress has been made in understanding the assumptions and limitations of the models used. A few examples include works such as \cite{2020ApJ...900...78D}, which show the importance of considering a wide range of theoretical models to assess the uncertainty of the measurements, and \cite{2021ApJ...920...88D}, which highlights the importance of remeasuring old data using the latest theoretical models. Works such as \cite{2021ApJ...907...31T} and \cite{2024ApJ...967...35L} test for possible deviations from the Kerr metric.  \cite{2020MNRAS.491..417R} and \cite{2021ApJ...913..129T} probe the effect of the geometric thickness of the accretion disk on reflection studies. \cite{2024ApJ...976..229M} and \cite{2025arXiv250111212H} analyzed the impact of returning radiation on the ability to properly quantify reflection features in X-ray spectra. \cite{2020arXiv201207469R} test the impact of the coronal geometry, by modeling it as a thin disk located along the rotation axis of the BH. Studies such as \cite{2018A&A...614A..44K, 2020ApJ...895...61R}, and \cite{2025MNRAS.536.2594L} test the accuracy of spin measurement techniques in current- and future-generation data. Furthermore, on the continuum fitting side, \cite{2024ApJ...962..101Z} argue that the spin measurements obtained for a few BH XBs are strongly model dependent. \cite{2024MNRAS.531..366M} probe the impact of emission of matter from within the plunging region of BH accretion disks on X-ray spectra, and \cite{2021MNRAS.500.3640S} show the importance of properly quantifying the uncertainties in the hardening factor on the ability to reliably constrain the BH spin through continuum fitting.

The studies of individual sources presented \cite{2020ApJ...900...78D, 2021ApJ...920...88D, 2023ApJ...947...39D, 2023ApJ...954...62D} serve as the basis on which a pipeline was built to reliably and uniformly measure the BH spin in multiple sources. This pipeline was presented in \cite{2023ApJ...946...19D}, where it was applied to multiple sources with archival observations, but with no prior relativistic reflection spin measurements. That work produced 10 new BH spin measurements. This work was further expanded in \cite{Draghis24}, where the entire NuSTAR archival data set of accreting BH XB systems was reanalyzed with the goal of constraining the BH spins in a uniform way. 

The work presented in \cite{Draghis24} compiled a sample of 36 spin measurements of accreting BHs in XB systems. These measurements were performed in a uniform and transparent way that properly encapsulates the systematic variability of the source between observations and the ability to characterize the spin regardless of the assumptions of our models. By considering the information obtained from all available observations through a Bayesian inference algorithm, we compiled the most complete set (as of the time of writing) of XB BH spin measurements. In the future, this data set will continue to expand through the discovery of new BH systems that undergo outbursts. This expanding dataset enables comparisons to the distribution of spins inferred based on gravitational-wave (GW) observations of merging  binary BHs (BBHs), studying the properties of the systems to which they belong in order to probe emerging trends regarding the physical characteristics of the sources, and also developing a more detailed understanding of the behavior of the models used in our analysis. 

\begin{figure}[ht]
    \centering
    \includegraphics[width= 0.45\textwidth]{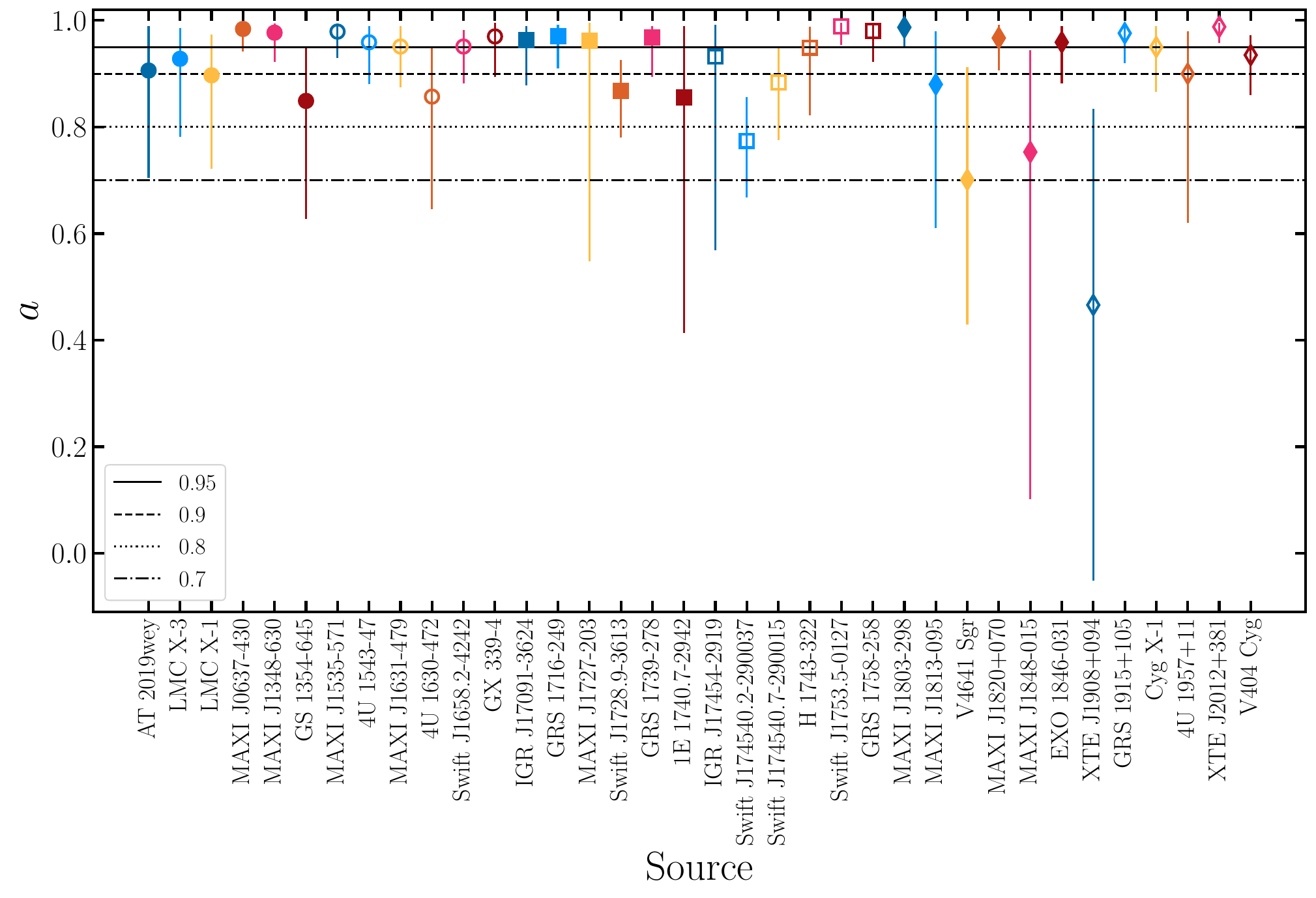}
    \caption{Illustration of the 36 spin measurements obtained throughout the analysis presented in \cite{Draghis24}. The horizontal lines represent spin values of $a=0.95$ (solid line), 0.9 (dashed line), 0.8 (dotted line), and 0.7 (dash-dotted line). The error bars represent the $1 \sigma$ uncertainties of the measurements, combining the statistical and systematic uncertainties associated with obtaining the measurements from multiple spectra of the same sources.}
    \label{fig:latest_spins}
\end{figure}

Figure \ref{fig:latest_spins} shows all the 36 measurements presented in \cite{Draghis24}. As highlighted in the article, $86\%$ of the measurements allow $a\geq0.95$, and $100\%$ allow $a\geq0.7$ (at the $1\;\sigma$ level). Furthermore, $28\%$ of the spins have a lower bound of their $1\;\sigma$ credible interval greater than $a\geq0.9$. At the lower end, only $\sim17\%$ of the measurements allow $a\leq0.6$ at the $1\;\sigma$ level. 

In \cite{Draghis24}, we focused on measuring the BH spin and inclination for 36 XB systems. Here, we continue the study of the same 36 XB systems. In particular, we focus on two main topics. First, we investigate the relation between the measured BH spin and other properties of the binary systems, with the goal of uncovering patterns in the observable properties of these systems that would enable studies of the formation and evolutionary pathways of BH XBs. Second, we explore the other parameters in the spectral fits in an attempt to uncover correlations that could influence the ability to quantify BH spin. The paper is structured as follows. In Section \ref{sec:data}, we describe the data set used in the analysis presented in this work. In Section \ref{sec:pop} we discuss the distribution of BH spins and the trends emerging when comparing the spin measurements to properties of the systems obtained from the literature. In Section \ref{sec:pars} we discuss possible parameter correlations and degeneracies in the spectral fitting. In Section \ref{sec:summary} we summarize the results of this work and suggest avenues for the future of the field of BH spin measurements.

\section{Data} \label{sec:data}

In the initial analysis of the 36 BH XBs that was presented in \cite{Draghis24}, we fit all existing NuSTAR spectra of the sources. We chose observations where the sources had an Eddington fraction in the range $10^{-3}\leq L/L_{\rm Edd}\leq 0.3$, assuming that the inner radius of the accretion disk reaches the ISCO. We fit all spectra using a baseline model, containing the contribution from a thermal accretion disk using the \texttt{diskbb} component, coronal emission modeled using the \texttt{powerlaw} component, and included Galactic absorption along the line of sight using the \texttt{TBabs} multiplicative component. We note that for spectra taken during particularly hard states where the disk contribution cannot be constrained given the low-energy NuSTAR bandpass limit of 3~keV, the parameters of the \texttt{diskbb} component converge to values of the disk temperature or normalization that are low enough so that the component does not significantly influence the total model. We advise future inference of trends that use the disk parameters to consider this effect and filter the data accordingly. Similarly, the absorption along the line of sight sometimes takes low values in systems where it cannot be constrained. This does not influence the quality of the fits.

After fitting the spectra using our baseline model, we refit the spectra using models that include the effects of relativistic reflection, through the \texttt{relxill} v.1.4.3 family of models (\citealt{2014MNRAS.444L.100D, 2014ApJ...782...76G}). This choice of version was made because it was the most recent version at the time when the study began, and we chose not to update to more recent versions of the model for the consistency of the analysis sample. We compared the fit statistic of the models that account for reflection with that produced by our baseline model to assess in which observations relativistic reflection is reliably detected in a statistically significant manner. We continued our analysis only for spectra in which reflection was clearly detected, even if the detection was weak and some parameters in the fit could not be reliably constrained. 

We fit all observations with a baseline array of six variations of the \texttt{relxill} model. All models account for absorption along the line of sight through the \texttt{TBabs} multiplicative component, and include a description of the thermal emission from the accretion disk through the \texttt{diskbb} component. The reflection is accounted for using six different variations of the \texttt{relxill} family of models. In \texttt{xspec} parlance, the models used are \texttt{TBabs * (diskbb + relxillX)}, with \texttt{relxillX} here indicating the different flavors of the model. First, we fit using the default \texttt{relxill} flavor, which is agnostic to the assumption of a coronal geometry, and parameterizes its emissivity onto the accretion disk as a broken power law, with the intensity proportional to the radius as $r^{-q_1}$ up to a radius of $R_{\rm br}$, followed at larger radii by an emissivity profile following $r^{-q_2}$. The second variation used adopts the \texttt{relxillCp} flavor, which describes the spectral emissivity profile of the corona as a Comptonization continuum, as opposed to \texttt{relxill} which describes it as a power law with a high-energy cutoff. The third variation of the reflection component uses the \texttt{relxilllp} flavor, which adopts a lamppost coronal geometry, where the compact corona is located at a height $h$ above the accretion disk, along the spin axis of the BH. We also explored the effect of variable disk density through the \texttt{relxillD} flavor, which also does not assume a coronal geometry, and describes the coronal spectral emissivity profile as a power law with a high-energy cutoff fixed at 300 keV. However, as the disk density could rarely be reliably constrained, we adopted three versions, with the density fixed at $\log(n/\rm cm^{-3})=15,\;17,$ and 19. Note that \cite{2018ApJ...855....3T} argued that the effects of disk density become important for $\log(n/\rm cm^{-3})\gtrsim19$, so we further explore the effect and limitations produced by this constraint in Section \ref{sec:density}. This set of six variations of the \texttt{relxill} model forms our baseline set of models. In sources where the residuals for the fits using the six variations indicated that additional components might be needed, we added a \texttt{gaussian} component to account for narrow absorption or emission features. In 36 of the total 245 spectra analyzed, the residuals presented additional evidence of complex obscuration. In those cases, we employed the \texttt{zxipcf} multiplicative component that models partial obscuration by an ionized absorber. We note that as the goal of this analysis is to constrain the broad reflection features, the choice of model for the ionized obscuration does not impact the robustness of our results, especially given the limited energy resolution of the NuSTAR spectra. In the particular case of V4641 Sgr, our best-fit models also include the \texttt{apec} component, as described in \cite{2023ApJ...946...19D}, which characterizes the emission from a collisionally ionized diffuse gas. For a complete explanation of the choice of models, see Section 2.1 in \cite{2023ApJ...946...19D}, and for a detailed explanation of the \texttt{relxill} family of models, their assumptions, and their parameters, see Section 3.1 in \cite{2021ApJ...920...88D} and Appendix A in \cite{2023ApJ...947...39D}.

After fitting all spectra with all models, we ran a Markov Chain Monte Carlo (MCMC) analysis on the best 1-3 models. For a description of the MCMC runs, see Section 2.2 in \cite{2023ApJ...946...19D}. We used the MCMC analysis to compute the deviance information criterion (DIC), both to ensure the convergence of the chains and to provide a statistically robust mean of selecting the ``best-performing" model. For the MCMC run with the lowest DIC, we computed the mode of the posterior distribution for all parameters, and the $\pm1\;\sigma$ credible intervals of the distribution. For all 245 NuSTAR spectra of the 36 sources analyzed in this sample, these numbers were compiled into a table. The table is available in Zenodo at \dataset[doi: 10.5281/zenodo.15801174]{https://doi.org/10.5281/zenodo.15801174}, and at the following \href{https://www.pdraghis.com/bh-xrbs-analysis}{link}\footnote{\url{https://www.pdraghis.com/bh-xrbs-analysis}}.

\section{Population study} \label{sec:pop}

\begin{deluxetable*}{l|cc|ccccc|c}
\tablecaption{Physical parameters of all the BHs in our sample.}
\label{tab:all_spins_and_masses}
\tablewidth{\textwidth} 
\tabletypesize{\scriptsize}
\tablehead{
\colhead{Source} & \colhead{$a$} & \colhead{$\theta_r\;[^\circ]$} & \colhead{$\rm M_{BH} \; [\rm M_\odot]$} & \colhead{$\rm d\;[\rm kpc]$} & \colhead{$\theta_i\;[^\circ]$} & \colhead{$\rm M_C \; [\rm M_\odot]$} & \colhead{$\rm P \;[\rm days]$} & \colhead{Reference}}
\startdata\\
AT 2019wey & $0.91_{-0.20}^{+0.08}$ & $14_{-10}^{+12}$ & \nodata & $5.5 \pm 4.5$ & \nodata & $\leq1$ & 0.666 & (1) \\
LMC X-3 & $0.93_{-0.15}^{+0.06}$ & $38_{-13}^{+14}$ & $6.98 \pm 0.56$ & $48.1 \pm 2.2$ & $69.24 \pm 0.72^\star$ & $3.63 \pm 0.57$ & 1.705 & (2) \\
LMC X-1 & $0.9_{-0.18}^{+0.08}$ & $50_{-13}^{+10}$ & $10.91 \pm 1.41$ & $48.1 \pm 2.2$ & $36.38 \pm 1.92^\star$ & $31.79 \pm 3.48$ & 3.909 & (3) \\
MAXI J0637-430 & $0.98_{-0.04}^{+0.01}$ & $63_{-10}^{+9}$ & \nodata & \nodata & \nodata & \nodata & \nodata & \nodata \\
MAXI J1348-630 & $0.98_{-0.06}^{+0.02}$ & $52_{-11}^{+8}$ & $11 \pm 2$ & $3.4 \pm 0.4$ & \nodata & \nodata & \nodata & (4) \\
GS 1354-645 & $0.8_{-0.2}^{+0.1}$ & $47_{-10}^{+11}$ & $7.6 \pm 0.7$ & $\geq25$ & $\leq79^\star$ & $0.91 \pm 0.32$ & 2.545 & (5) \\
MAXI J1535-571 & $0.98_{-0.05}^{+0.02}$ & $44_{-19}^{+17}$ & $8.9 \pm 1.0$ & $4.1 \pm 0.6$ & \nodata & \nodata & \nodata & (6,7) \\
4U 1543-47 & $0.96_{-0.08}^{+0.03}$ & $67_{-8}^{+7}$ & $9.4 \pm 2.0$ & $7.5 \pm 1$ & \nodata & \nodata & 1.123 & (8,9) \\
MAXI J1631-479 & $0.95_{-0.08}^{+0.04}$ & $22_{-12}^{+10}$ & \nodata & \nodata & \nodata & \nodata & \nodata & \nodata \\
4U 1630-472 & $0.86_{-0.21}^{+0.10}$ & $55_{-11}^{+8}$ & \nodata & \nodata & \nodata & \nodata & \nodata & \nodata \\
Swift J1658.2-4242 & $0.95_{-0.07}^{+0.03}$ & $50_{-10}^{+9}$ & \nodata & \nodata & \nodata & \nodata & \nodata & \nodata \\
GX 339-4 & $0.97_{-0.08}^{+0.03}$ & $49 \pm 14$ & $9.0 \pm 1.5$ & $8.4 \pm 0.8$ & $57.5 \pm 20.5^\star$ & $1.6 \pm 0.5$ & 1.759 & (10,11) \\
IGR J17091-3624 & $0.96_{-0.08}^{+0.03}$ & $47_{-11}^{+10}$ & $12.15 \pm 3.45$ & $12.6 \pm 2.0$ & \nodata & \nodata & \nodata & (12) \\
GRS 1716-249 & $0.97_{-0.06}^{+0.02}$ & $59_{-12}^{+7}$ & $6.4_{-2.0}^{+3.2}$ & $6.9 \pm 1.1$ & $61 \pm 15^\star$ & $\leq0.69$ & 0.278 & (13) \\
MAXI J1727-203 & $0.96_{-0.41}^{+0.03}$ & $65_{-14}^{+11}$ & \nodata & \nodata & \nodata & \nodata & \nodata & \nodata \\
Swift J1728.9-3613 & $0.87_{-0.09}^{+0.06}$ & $7_{-3}^{+8}$ & \nodata & \nodata & \nodata & \nodata & \nodata & \nodata \\
GRS 1739-278 & $0.97_{-0.07}^{+0.02}$ & $70_{-11}^{+5}$ & $6.75 \pm 2.75$ & $7.25 \pm 1.25$ & \nodata & \nodata & \nodata & (14,15) \\
1E 1740.7-2942 & $0.9_{-0.4}^{+0.1}$ & $31_{-18}^{+29}$ & $5.0 \pm 0.5$ & $8.5 \pm 2.0$ & \nodata & \nodata & 12.730 & (16,17) \\
IGR J17454-2919 & $0.93_{-0.36}^{+0.06}$ & $54_{-14}^{+15}$ & \nodata & \nodata & \nodata & \nodata & \nodata & \nodata \\
Swift J174540.2-290037 & $0.77_{-0.11}^{+0.08}$ & $31_{-9}^{+8}$ & \nodata & \nodata & \nodata & \nodata & \nodata & \nodata \\
Swift J174540.7-290015 & $0.88_{-0.11}^{+0.07}$ & $63_{-8}^{+10}$ & \nodata & \nodata & \nodata & \nodata & \nodata & \nodata \\
H 1743-322 & $0.95_{-0.13}^{+0.04}$ & $54_{-13}^{+12}$ & $11.21_{-1.96}^{+1.65}$ & $8.5 \pm 0.8$ & $75.0 \pm 3.0^\dagger$ & \nodata & \nodata & (18,19) \\
Swift J1753.5-0127 & $0.989_{-0.035}^{+0.007}$ & $73 \pm 8$ & $8.8 \pm 1.3$ & $3.9 \pm 0.7$ & $79 \pm 5^\star$ & $0.2 \pm 0.06$ & 0.136 & (20) \\
GRS 1758-258 & $0.98_{-0.06}^{+0.01}$ & $67_{-13}^{+8}$ & \nodata & \nodata & \nodata & \nodata & 18.450 & (21) \\
MAXI J1803-298 & $0.987_{-0.037}^{+0.007}$ & $72_{-9}^{+6}$ & $8.0 \pm 4.5$ & \nodata & $67.5 \pm 7.5^\Diamond$ & $0.77 \pm 0.05$ & 0.292 & (22) \\
MAXI J1813-095 & $0.9_{-0.3}^{+0.1}$ & $42_{-13}^{+11}$ & $7.4 \pm 1.5$ & $6 \pm 1$ & \nodata & \nodata & \nodata & (23) \\
V4641 Sgr & $0.7_{-0.3}^{+0.2}$ & $66_{-11}^{+7}$ & $6.4 \pm 0.6$ & $6.2 \pm 0.7$ & $72.3 \pm 4.1^\star$ & $2.9 \pm 0.4$ & 2.817 & (24,25) \\
MAXI J1820+070 & $0.97_{-0.06}^{+0.02}$ & $64_{-9}^{+8}$ & $8.48 \pm 0.79$ & $2.96 \pm 0.33$ & $63 \pm 3^\dagger$ & $0.61 \pm 0.13$ & 0.703 & (26,27) \\
MAXI J1848-015 & $0.8_{-0.7}^{+0.2}$ & $29_{-10}^{+13}$ & \nodata & \nodata & \nodata & \nodata & \nodata & \nodata \\
EXO 1846-031 & $0.96_{-0.08}^{+0.03}$ & $62_{-10}^{+9}$ & \nodata & \nodata & \nodata & \nodata & \nodata & \nodata \\
XTE J1908+094 & $0.5_{-0.5}^{+0.4}$ & $28 \pm 11$ & \nodata & $6.5 \pm 3.5$ & $\geq79^\dagger$ & \nodata & \nodata & (28,29) \\
GRS 1915+105 & $0.98_{-0.06}^{+0.02}$ & $60 \pm 8$ & $12.4 \pm 0.2$ & $8.6_{-1.8}^{+2.0}$ & $60 \pm 5^\dagger$ & $0.47 \pm 0.27$ & 33.833 & (30,31) \\
Cyg X-1 & $0.95_{-0.08}^{+0.04}$ & $47_{-11}^{+9}$ & $21.2 \pm 2.2$ & $2.22 \pm 0.18$ & $27.6 \pm 0.7^\star$ & $40.9 \pm 7.4$ & 5.600 & (32,33) \\
4U 1957+11 & $0.9_{-0.28}^{+0.08}$ & $52_{-13}^{+12}$ & $5 \pm 1$ & $7.5 \pm 2.5$ & \nodata & $\leq1$ & 0.389 & (34,35,36) \\
XTE J2012+381 & $0.988_{-0.030}^{+0.008}$ & $68_{-11}^{+6}$ & \nodata & \nodata & \nodata & \nodata & \nodata & \nodata \\
V404 Cyg & $0.94_{-0.08}^{+0.04}$ & $37_{-8}^{+9}$ & $12 \pm 3$ & $2.39 \pm 0.14$ & $67_{-1}^{+3}$$^\star$ & $0.72 \pm 0.19$ & 6.471 & (37,38,39,40) \\
\enddata

\tablecomments{List of references: 1:\cite{2021ApJ...920..120Y}; 2:\cite{2014ApJ...794..154O}; 3:\cite{2009ApJ...697..573O}; 4:\cite{2021AA...647A...7L}; 5:\cite{2009ApJS..181..238C}; 6:\cite{2019ApJ...875....4S}; 7:\cite{2019MNRAS.488L.129C}; 8:\cite{2004ApJ...610..378P}; 9:\cite{1998ApJ...499..375O}; 10:\cite{2017ApJ...846..132H}; 11:\cite{2016ApJ...821L...6P}; 12:\cite{2015ApJ...807..108I}; 13:\cite{2023MNRAS.526.5209C}; 14:\cite{2018PASJ...70...67W}; 15:\cite{1996AA...314L..21G}; 16:\cite{2002ApJ...578L.129S}; 17:\cite{2020MNRAS.493.2694S}; 18:\cite{2012ApJ...745L...7S}; 19:\cite{2017ApJ...834...88M}; 20:\cite{2025AA...694A.119Y}; 21:\cite{2002ApJ...578L.129S}; 22:\cite{2022MNRAS.511.3922J}; 23:\cite{2021RAA....21..125J}; 24:\cite{2001ApJ...555..489O}; 25:\cite{2014ApJ...784....2M}; 26:\cite{2020MNRAS.493L..81A}; 27:\cite{2020ApJ...893L..37T}; 28:\cite{2017MNRAS.468.2788R}; 29:\cite{2006MNRAS.365.1387C}; 30:\cite{2013ApJ...768..185S}; 31:\cite{2014ApJ...796....2R}; 32:\cite{1999AA...343..861B}; 33:\cite{2021Sci...371.1046M}; 34:\cite{1987ApJ...312..739T}; 35:\cite{2023ApJ...944..165B}; 36:\cite{2021RAA....21..214S}; 37:\cite{2010ApJ...716.1105K}; 38:\cite{1994MNRAS.271L..10S}; 39:\cite{2019MNRAS.488.1356C}; 40:\cite{2009ApJ...706L.230M}. $^\star$ indicates a binary inclination measurement; $^\dagger$ represents an inference of inclination based on radio jets; $^\Diamond$ represents an inclination estimate based on the presence or absence of dips in X-ray light curves.}
\end{deluxetable*}

\begin{figure*}[!ht]
    \centering
    \includegraphics[width= 0.95\textwidth]{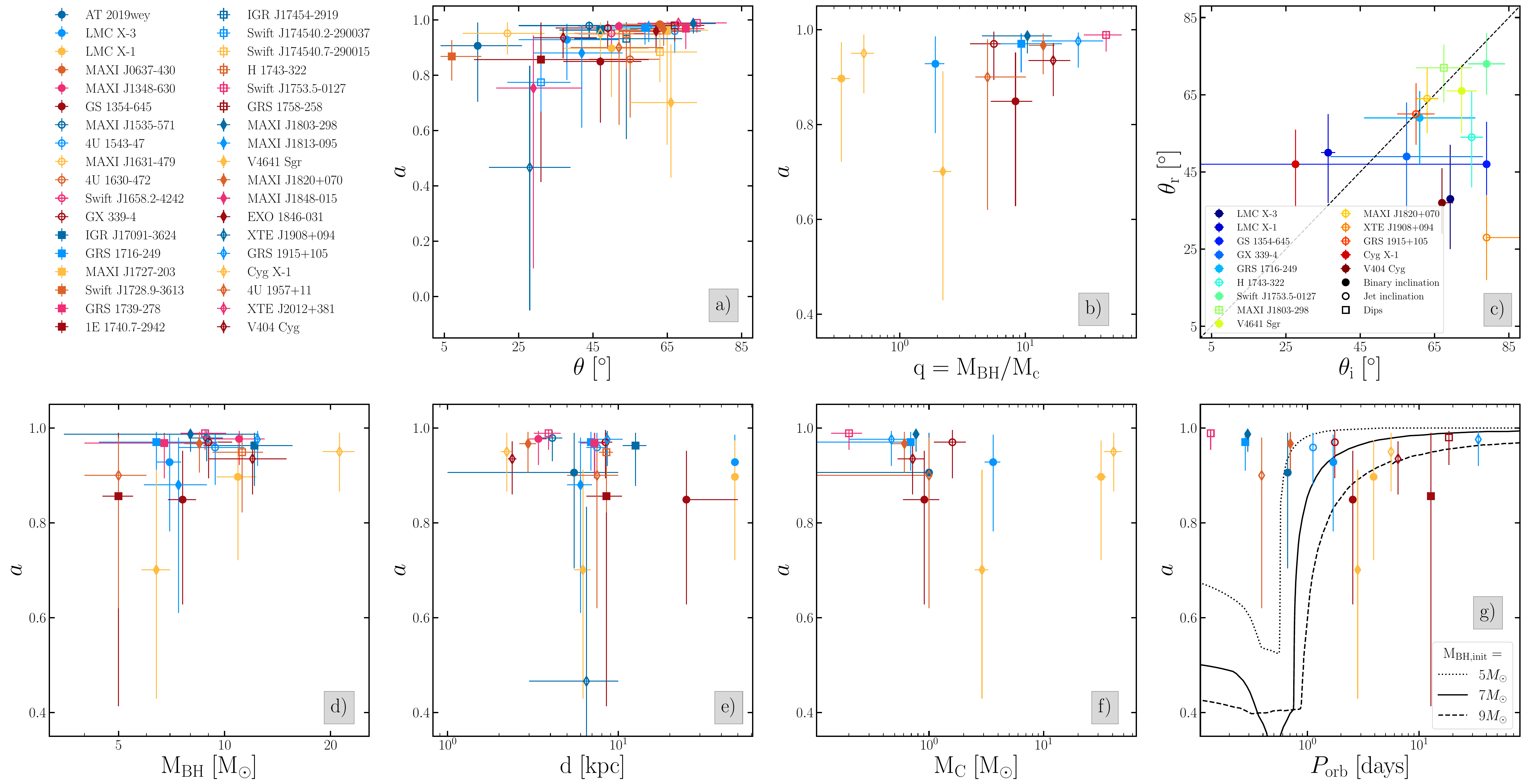}
    \caption{The legend in the top-left panel shows the markers that were used in the all panels except for the top right to identify the 36 sources analyzed in this work, including in Figure \ref{fig:latest_spins}. Top row, from left to right: (a) spin measurements vs. inclination of the inner regions of the accretion disk for the 36 sources in our sample, as measured in \cite{Draghis24}. (b) BH spin vs. ratio of BH mass to stellar companion mass. (c) Inclination measured through reflection vs. inclination measured through independent means, reported in the literature. Solid circles represent binary inclinations, the empty circles represent measurements based on radio jet morphology, and the empty square represents an inclination inference based on the presence of dips in the X-ray light curves. Bottom row, from left to right, show the BH spin vs BH mass (panel d), distance to the system (panel e), stellar companion mass (panel f), and orbital period of the systems analyzed in this work (panel g). The curves in the lower-right panel represent the theoretical predictions of the maximum spin that a BH in an LMXB can achieve for a given orbital period, as a function of the initial BH mass, with the dotted line representing the values for a $5\;M_\odot$ BH, the solid line showing a $7\;M_\odot$ BH, and the dashed line showing a $9\;M_\odot$ BH, as computed by \cite{2015ApJ...800...17F}. Note that the panels only show points where where values exist in the literature for the different measurements. }
    \label{fig:spin_vs_all}
\end{figure*}

Table \ref{tab:all_spins_and_masses} presents a few physical properties of the 36 BH systems treated in this work, obtained from a review of the literature of the sources. The table includes the BH spin and inclination of the inner accretion disk, as measured in \cite{Draghis24}, the mass of the BH ($M_{\rm BH}$), the distance to the system ($d$), the inclination obtained through independent measurements, not related to X-ray spectral analysis ($\theta_i$), the mass of the stellar companion ($M_{\rm C}$), and the orbital period of the binary system ($P$), together with references to the work that reported these measurements. We note that the table only contains the values for sources where they exist in the literature. Using the values presented in Table \ref{tab:all_spins_and_masses}, we present the relation between a few interesting combinations of parameters, shown in Figure \ref{fig:spin_vs_all}. 

\begin{figure}[ht!]
    \centering
    \includegraphics[width= 0.48\textwidth]{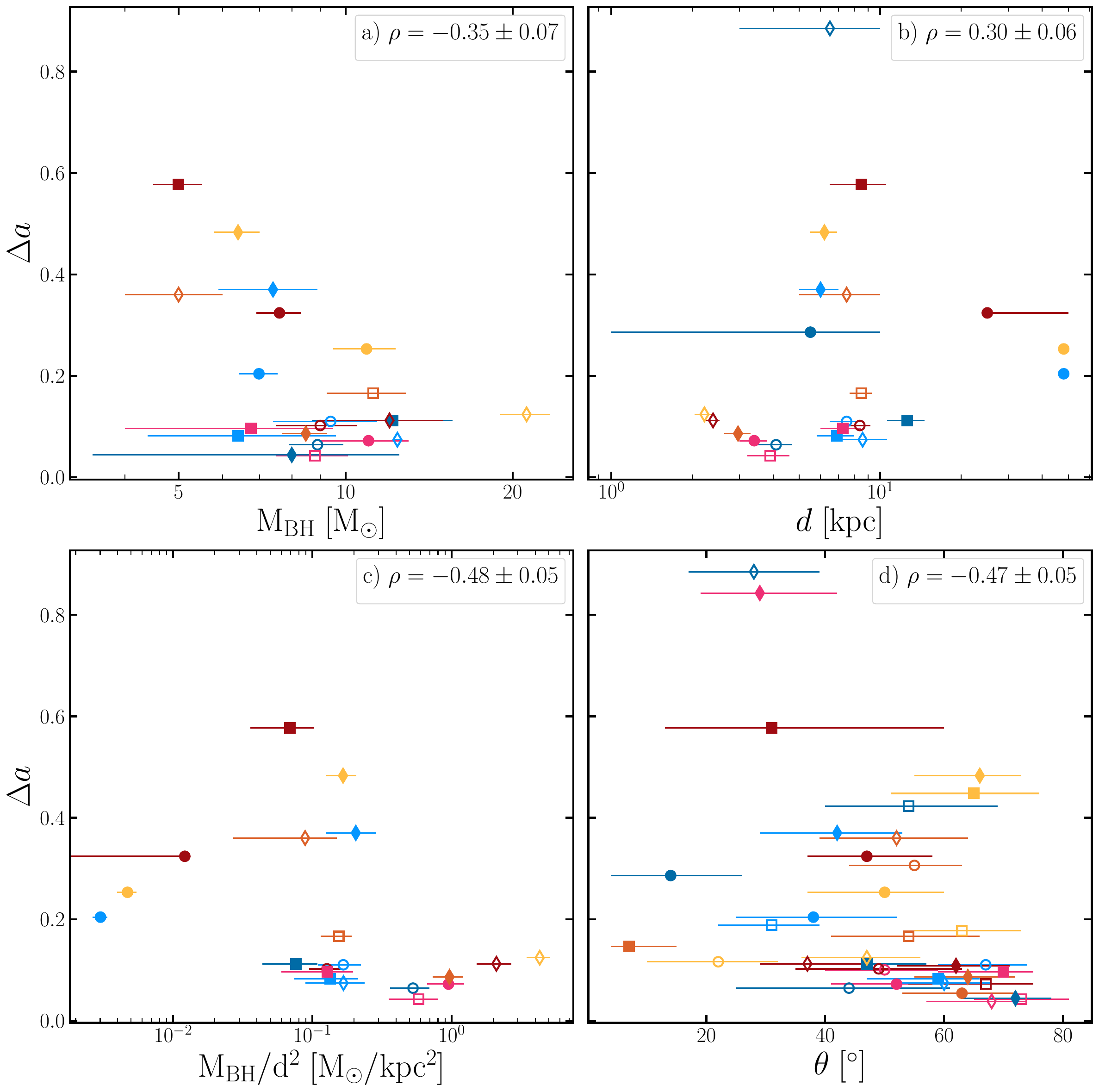}
    \caption{Uncertainty in spin measurements vs. BH mass (panel a), distance to the system (panel b), the ratio of BH mass to distance to the system squared (panel c), and the measured inclination (panel d). The $\rho$ values in the individual panels represent the Spearman correlation coefficients for the samples.}
    \label{fig:da_vs_parameters}
\end{figure}

Panel (a) in Figure \ref{fig:spin_vs_all} shows the BH spins measured by reflection versus the inclination of the inner disk inferred by reflection. This trend is particularly interesting, as both the spin and the viewing inclination influence the shape of the Fe line in the reflected spectrum by broadening it. However, the BH spin preferentially influences the red wing of the line, while the inclination is most important on the blue wing of the line. The distribution of measured inclinations peaks around $60^\circ$, consistent with isotropic orientations of the systems. Although no trend is clearly present across the sample of 36 sources, it does appear visually as if systems with low inclinations (i.e. $\theta \leq 40^\circ$) seem to be more slowly rotating than the many nearly maximally spinning BHs that are seen to have higher inclinations, and the sizes of the uncertainties of the measurements are generally larger. This potentially indicates that at smaller inclinations the spins are more difficult to constrain. This trend is indeed supported by the Spearman correlation coefficient resulting from comparing the uncertainty in the measured spin with the measured inclination (Figure \ref{fig:da_vs_parameters}, panel d), with a value of $\rho=-0.47\pm0.05$, indicating a moderate negative correlation.

Panel (b) in Figure \ref{fig:spin_vs_all} shows the BH spin versus the ratio of the mass of the BH in the system to the mass of the binary companion. It is important to note that in making this figure, only a few sources have estimates of both the BH and companion masses, so emerging trends should be interpreted while also considering observational selection effects that prevent reliably constraining the masses of the components. The interesting pattern emerging from this plot is that higher mass ratios, translating to particularly heavy BHs being orbited by low-mass stellar companions, allow more precise spin measurements. Simultaneously, an interesting point emerges when inspecting panel (f) in Figure \ref{fig:spin_vs_all}: the number of systems with subsolar-mass companions drastically outnumbers systems with companions having a mass of a few solar masses.

Panel (c) in Figure \ref{fig:spin_vs_all} shows the relationship between the inclinations measured by relativistic reflection and the inclinations inferred by independent means. The solid circles represent inclinations inferred based on the light curves of the binary system, the empty circles represent inclinations inferred based on the morphology of the observed radio jets, and the empty square points indicate inference on inclination placed based on dips being present in the X-ray light curves. Although the uncertainties of the X-ray inclination measurements are often large, the measurements are generally in good agreement. 

The bottom row in Figure \ref{fig:spin_vs_all} shows the spin of the BH, inferred using relativistic reflection, versus the mass of the BH (panel d), the distance to the system (panel e), the companion mass (panel f) and the orbital period of the binary system (panel g). 
We note that no trends between the measured spin values and other parameters are clear, but a few trends are apparent when analyzing the uncertainty of the spins in relation to the other system properties. When examining the influence of the BH mass on the spin, perhaps the most noticeable trend is the increase in measurement precision with increasing BH mass. Panel (a) in Figure \ref{fig:da_vs_parameters} shows the uncertainty of the spins measured in this sample compared to the masses of the BHs. The Spearman correlation test returns a correlation coefficient of $\rho=-0.35\pm0.07$, suggesting a moderate negative trend, with the spin uncertainty decreasing with increasing BH mass. This trend may be due to higher-mass BHs having higher accretion rates and, therefore, being brighter and producing spectra with better signal-to-noise ratios (SNRs). However, it is important to note that the observed flux is also proportional to the inverse of the distance to the system squared, potentially strongly contributing to any trends inferred based on this small sample.

The panel indicating the dependence of the BH spins on the distance to the observed systems (e) shows trends that behave as expected: the uncertainty of the spin measurements increases with distance, followed by a drop in detectability at distances larger than $\sim 10\;\rm kpc$, suggested by a reduction in the number of known systems at large distances. The exceptions are the two extragalactic sources in our sample (LMC X-1 and LMC X-3), and GS 1354-645, for which the distance to the system is still an active topic of debate. Panel (b) in Figure \ref{fig:da_vs_parameters} shows the uncertainty in the measured BH spin versus the distances to the systems. The Spearman correlation coefficient of $\rho=0.30\pm0.06$ suggests a mild positive correlation between the spin uncertainty and increasing distance.

To constrain the joint impact of the BH mass and distance to the system on the ability to precisely constrain BH spin, in panel (c) of Figure \ref{fig:da_vs_parameters} we plot the uncertainty in the measured BH spins vs. the ratio of the BH mass to the square of the distance to the systems. This choice was made because the observed fluxes will be proportional to the mass accretion rate, which is proportional to the BH mass. At the same time, the observed flux will be inversely proportional to the square of the distance to the systems. When combining the mild correlations highlighted in panels (a) and (b), the Spearman correlation coefficient for the sample in panel (c) produces a stronger negative correlation. Intuitively, this trend is expected and suggests that when BHs are more massive and closer to us, the measured spins will be more precise.

Panel (g) in Figure \ref{fig:spin_vs_all} shows the spins of the BHs in our sample versus the orbital periods of the binary systems. By comparing this figure with the theoretical predictions of \cite{2015ApJ...800...17F} for the maximum spin that a BH in an LMXB can have if entirely acquired through accretion (shown through the different lines for different initial BH masses), the high spins seen for systems with short orbital periods indicate that the BH rotation could not have been obtained solely by accretion. 

\subsection{Spin Distribution}\label{sec:distributions}
\begin{figure}[ht]
    \centering
    \includegraphics[width= 0.48\textwidth]{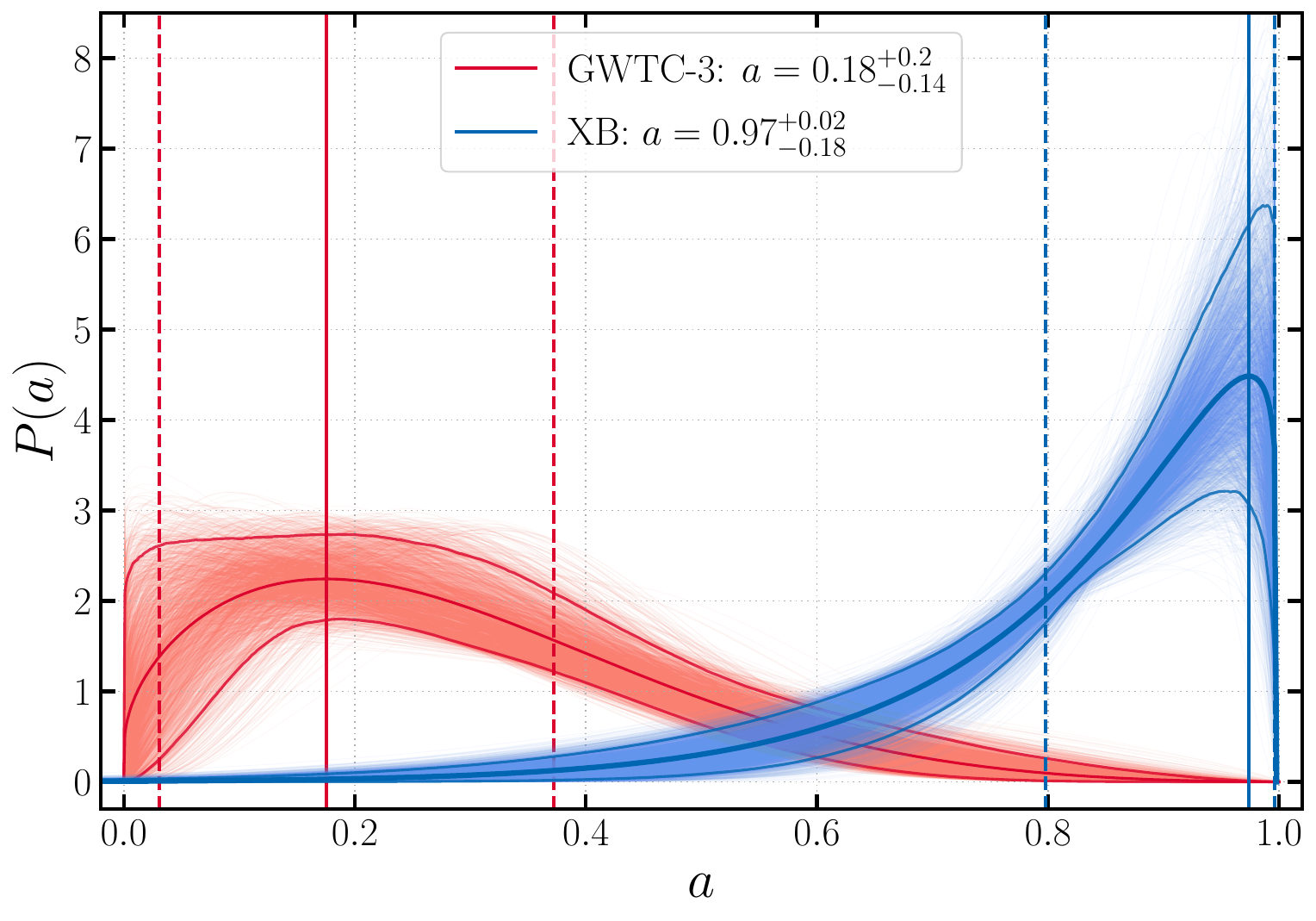}
    \caption{Probability distributions for the distributions of BH spins inferred based on GWTC-3 (red) and the measurements in this work (blue). The solid curves represent the mean and central 90\% credible bounds inferred on the distributions. The thin lines represent individual draws emerging from the Bayesian analysis used to infer the distributions. The vertical solid and dashed lines represent the mode and $\pm1\;\sigma$ credible intervals of the distributions, and the values are shown in the figure legend.}
    \label{fig:spin_distributions_final}
\end{figure}

Similarly to the analysis performed in \cite{2023PhRvX..13a1048A} on spin measurements obtained through GW data, we performed a Bayesian inference with the aim of characterizing the observed distribution of BH spins in our sample. The solid blue and red curves in Figure \ref{fig:spin_distributions_final} show the mean and central 90\% credible bounds inferred on the probability density functions $P(a)$ for the spins of BHs in XBs (blue) and in BBHs (red). The light blue and red lines represent individual draws from the posterior samples resulting from the Bayesian inference algorithm that characterizes the two spin distributions. The vertical solid and dashed lines represent the modes and $\pm1\;\sigma$ of the mean distributions (central solid lines) for the two distinct populations of BHs. We find that the median distribution of spins in XBs (central solid blue line in Figure \ref{fig:spin_distributions_final}) is well approximated by a beta distribution, with coefficients $\alpha=5.66$ and $\beta=1.09$:
\begin{equation}
\begin{aligned}
    P(a) &=\textrm{constant} \times a^{\alpha-1}(1-a)^{\beta-1}\\
    &=\frac{\Gamma(\alpha+\beta)}{\Gamma(\alpha)\Gamma(\beta)}\times a^{\alpha-1}(1-a)^{\beta-1} 
\end{aligned}
\end{equation}
Here, $\Gamma(z)=\int_0^{\infty}t^{z-1}r^{-t}dt$ is the gamma function.

The two distributions are clearly distinct, with the spin distribution inferred for the BHs in BBH systems detected through GW preferring low spins, and the XB BHs preferentially spinning nearly maximally. It is important to note that the distribution shown in red, based on the results of GWTC-3, does account for observational selection effects, while the distribution shown in blue, based on the results of the work presented in \cite{Draghis24}, only shows the \textit{observed} spin distribution of the BHs in XBs. This experiment expands on the findings of \cite{2022ApJ...929L..26F}. In the future, properly understanding the selection biases that go into producing the observed sample of BH XBs is crucial for understanding the joint population of stellar-mass BHs. 

\subsection{Implications}\label{sec:implications}

Some of the trends highlighted by Figures \ref{fig:spin_vs_all} and \ref{fig:da_vs_parameters} confirm past literature results, while others open possible new directions for the field of study of BH XBs. In this section, we discuss the implications of the results of this study of the 36 BH XBs in our sample, when considering the system properties in relation to the BH spins. 

The masses of BHs in XBs tend to be lower than $\sim20\;\rm M_\odot$, as found by \cite{2016A&A...587A..61C}. Works such as \cite{2021ApJ...921..131J} find that the distribution of observed Galactic LMXBs is biased against the most massive BHs. The highest BH mass in this limited sample is around $\sim 20 \; \rm M_\odot$ (Cygnus X-1), significantly lower than the majority of the BHs observed to merge through GWs. However, when considering the observational biases associated with GW detection, the distribution of BH masses is still peaked at low values and decreases with increasing mass (\citealt{2023PhRvX..13a1048A}). Nevertheless, the recent discovery of a $33\;\rm M_\odot$ BH (\citealt{2024arXiv240410486G}) suggests that more massive BHs do exist within our Galaxy, despite not being observed through X-rays. Our analysis indicates that the uncertainty in BH spin measurement decreases with an increase in BH mass. This tentative trend is likely driven by the impact of the BH mass on the overall flux of the system, with higher-mass BHs accreting at the same Eddington fraction being more luminous, and therefore producing spectra with higher SNRs.

However, it is important to also factor in the limited sample size and the potential that the SNRs are impacted by the distances to the sources. In our analysis, this trend persists when factoring in the distance to the systems. This analysis will be further expanded in the future, as more sources are detected, and more accurate masses and distances are measured. In particular, works such as \cite{2024MNRAS.530..424A} aim to provide reliable methods for estimating distances to Galactic XBs based on the X-ray thermal continuum. In the future, such tools could help enhance our understanding of the locations of XBs in the Galaxy, leading to an improved understanding of the observational selection effects that inhibit our ability to discover new sources. Additionally, while it is informative to treat all systems in a uniform way, it is important to note that each BH system has its own particularities and that even different outbursts of the same source can look very different.

When analyzing the inclinations measured through relativistic reflection in relation to those measured through independent means, a few measurements do not show agreement. This could perhaps indicate a difference between the inclinations of the innermost regions of the accretion disk (inferred through reflection) and the binary inclination. At the same time, it is worth noting that sources such as V404 Cygni exhibit jet precession (\citealt{2019Natur.569..374M}), which could influence the inferred jet inclinations. Given that the angular momentum vector of the inner accretion disk is expected to be aligned to the spin of the BH, this would perhaps indicate the importance of natal kicks in misaligning the BHs compared to their binary orbits. This result would have important implications on the ability to disentangle spin magnitudes from spin orientations inferred based on GW measurements of BBHs, in the assumption of isolated formation of the merging systems. 

In comparing the pairs of BH spin and orbital period of the binary, the most important observation is the presence of systems with spins higher than the theoretical predictions of \cite{2015ApJ...800...17F} allow for a binary of a given period. These predictions assume that the BH angular momentum is acquired only through accretion. This indicates that while accretion can spin up BHs, at least a large part of the angular momentum of the BH must have been present at birth, confirming the findings presented in \cite{2023ApJ...947...39D}, who measured the spin of the newborn BH Swift J1728.9-3613. Note that our sample contains more data points compared to those of \cite{2015ApJ...800...17F}, especially at low orbital periods. This highlights the importance of characterizing sources with as much variety in the system properties as possible. Concurrently, comparing the data set in this work to that of \cite{2015ApJ...800...17F} again highlights the implications of the lack of low spin measurements on understanding the formation and evolution channels of stellar-mass BHs. Future theoretical models describing supernovae and the formation and evolution of BHs in binaries must be able to reproduce the observed rapidly rotating BHs prior to any episodes of accretion. 

Given the orbital periods at which these systems are observed, nearly all donor stars must have evolved past the main sequence to be able to fill their Roche lobes given the orbital separations. First, this suggests that the XB phase must be relatively short-lived, potentially explaining the relatively low number of accreting XBs that we see in our Galaxy compared to theoretical expectations of the total number of BHs in the Galaxy. Second, the progenitor star that produced the BH must have been a massive star. Given that the companions generally have low masses, this raises the question of how systems with such extreme mass ratios are formed. This has possible implications regarding the observational selection effects of BH binaries and could imply that systems with similar component masses might not enter an XB phase. Such mass ratios, at low orbital periods, could lead to common envelope phases, where the companion star loses drastic amounts of mass, resulting in the observed low companion masses. On the other hand, when the masses of the components are similar, at high orbital periods, they will rarely enter a Roche lobe overflow phase because of the large separation, and the only such systems that we see are the few, very short-lived HMXBs. For large mass ratios, small-mass companions may not survive the post-main-sequence evolution of the massive BH progenitor, whereas systems with large orbital periods will only enter the XB phase once the companion evolves past the main sequence. Although qualitative, this reasoning motivates future studies that aim to explain the population of observed BH binaries.


The discovery of GWs enabled characterization of the spin of the components merging in BBH systems. However, the spins inferred from these events differ from those measured in well-studied XBs, with GW results indicating a distribution that prefers low spins (\citealt{2023PhRvX..13a1048A}), while the BHs in XBs are almost always observed to be rapidly rotating (\citealt{2023ApJ...946...19D}). Furthermore, the masses of BHs in BBH systems are generally larger than of those in XBs. The spins inferred for the BHs in BBHs are likely to be influenced by prior modeling assumptions (\citealt{2017PhRvL.119y1103V}), while the spin measurements of the BHs in XBs often underestimate systematic uncertainties. This suggests that at least part of the apparent discrepancy between the two distinct spin distributions could be artificial. However, the discrepancy could also have physical origins. For example, part of the angular momentum of the BH could be obtained through accretion. However, it is unlikely that the BHs in the short-lived HMXBs (those that have massive stellar companions) have had enough time to accrete enough mass to drastically change their spin (\citealt{1995ApJS..100..217I}), while the BHs in LMXBs (those that have small stellar companions) likely do not have access to a large enough matter reservoir to change their spin significantly. As an example, a $10\;M_\odot$ BH must accrete $\sim6.8\;M_\odot$ to increase its spin from $a=0$ to $a=0.9$, and $\sim5\;M_\odot$ to increase its spin from $a=0.9$ to $a=0.998$. For the same $10\;M_\odot$ BH, accreting $1\;M_\odot$ would increase its spin from $a=0$ to $a\sim0.3$, or from $a=0.9$ to $a=0.94$. The high spins of BHs in XBs must be close to their natal values, thereby providing crucial clues to how they formed. Thus, robustly measuring the differences in spin distributions between the BBHs observed in GW mergers and the BHs in XBs is the key to understanding whether the spin distribution discrepancy between GW and X-ray measured spins is artificial, or a consequence of different formation mechanisms. 

Two points emerge from this conclusion. First, why are there no slowly spinning BHs measured with reflection? This point is particularly interesting, as some sources in our analysis have spin measurements using continuum fitting that predict low spins (see, e.g. \citealt{2020MNRAS.493.5389F} for MAXI J1820+070). Relativistic reflection was shown to be able to produce intermediate spin measurements in AGN (see, e.g. \citealt{2022MNRAS.514.2568S}), and a naive expectation is that lower spins would be easier to quantify, as the shape of the Fe line would be less broadened, and therefore easier to distinguish from the underlying continuum. However, this question persists, and future planned studies aim to answer this issue. Second, what are the selection effects that impact the observed distribution of spins in BH XBs? Similarly, this is an extremely intricate issue that probably requires careful population synthesis studies, and works such as \cite{2024MNRAS.534.1868G} showcase the power of such simulations to explain the formation processes of binaries containing BHs. 

\section{Parameter Correlations} \label{sec:pars}

Through the analysis presented in \cite{Draghis24}, 245 NuSTAR spectra of 36 BH XBs were analyzed. Although the main focus of that work was to characterize the spins of the 36 BHs, a by-product of the analysis is a large data set that contains all the parameters of the spectral fitting. In this section, we focus on an analysis of this entire data set of spectral fits. This data set is built using uniform assumptions regarding the data and the models. Using this dataset, we can begin to explore the impact of model behavior on spin measurements in a source-independent way, and explore possible hidden correlations and degeneracies between parameters on a global scale, which the analysis of a few observations of one source would not immediately reveal. 

\begin{figure}[ht!]
    \centering
    \includegraphics[width= 0.48\textwidth]{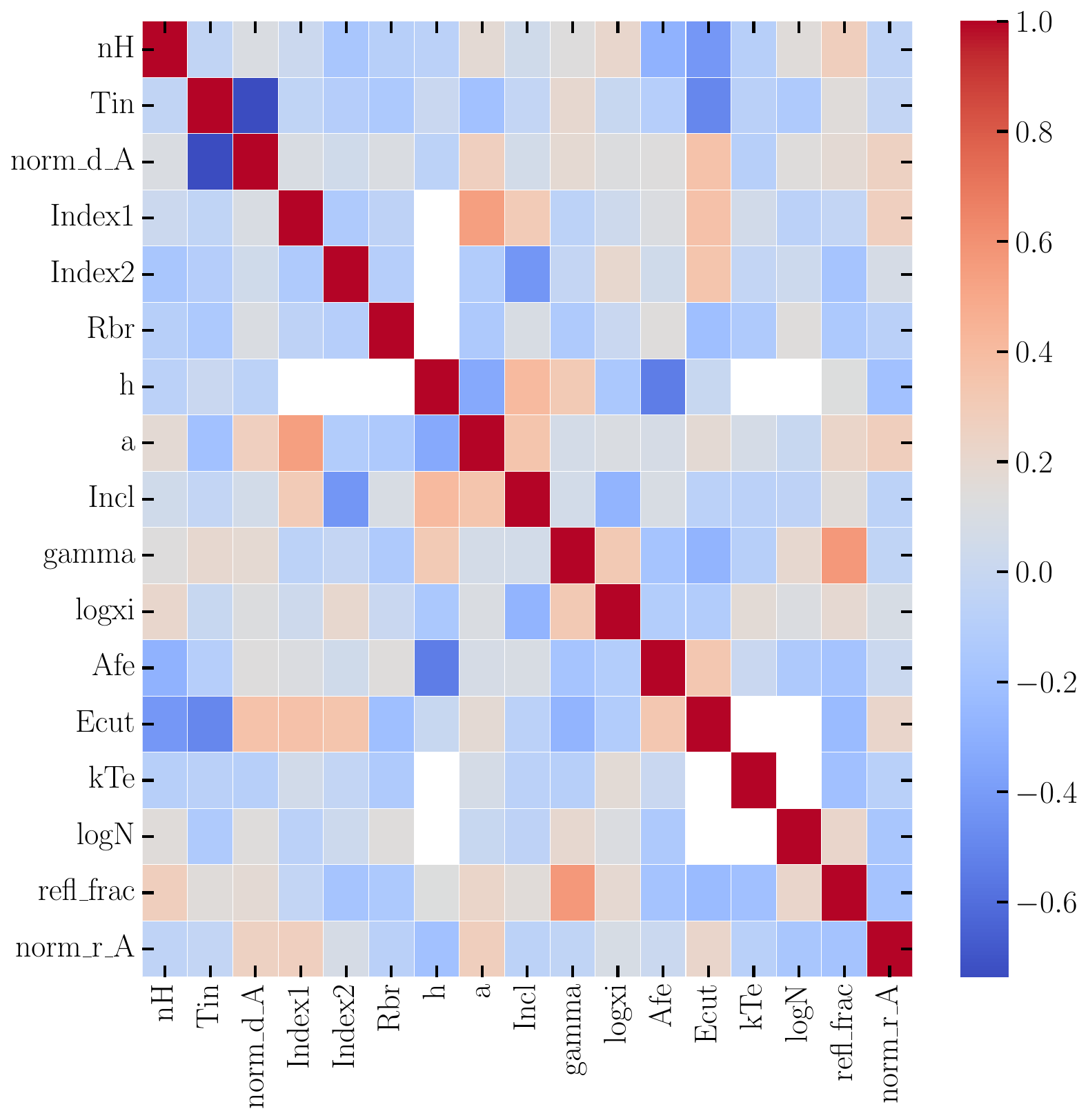}
    \caption{Correlation matrix for parameters of interest. Redder colors represent stronger positive correlations, while bluer colors represent stronger negative correlations. See Appendix \ref{sec:full_corr_matrix} for the full correlation matrix, together with the numerical values of the correlation coefficients and an explanation for how these were computed. The labels of the parameters represent the names extracted from the models in \texttt{xspec}.}
    \label{fig:correlation_matrix}
\end{figure}

Figure \ref{fig:correlation_matrix} shows a correlation matrix between some parameters of interest in the models used to fit all the spectra in our analysis. Redder colors indicate stronger positive correlations, while bluer colors indicate stronger negative correlations. The white entries represent combinations of parameters that do not occur in the models, e.g. coronal height $h$ and emissivity indices $q_{1,2}$, highlighted in the figure through the labels ``Index1" and ``Index2." The complete correlation matrix, together with the numerically expressed correlation coefficients, is presented in Appendix \ref{sec:full_corr_matrix}. This figure suggests parameter combinations that show potentially important parameter correlations and inspires the exploration pursued in the rest of this paper. Using the correlation coefficients presented in this figure, we visually inspect the combinations of parameters that show stronger trends, and in Sections \ref{sec:spin_corr} and \ref{sec:other_corr} we highlight some interesting results of this analysis. This figure has the potential to inspire future exploration of the parameter space.

\subsection{Correlations with Spin} \label{sec:spin_corr}

\begin{figure*}[ht!]
    \centering
    \includegraphics[width= 0.85\textwidth]{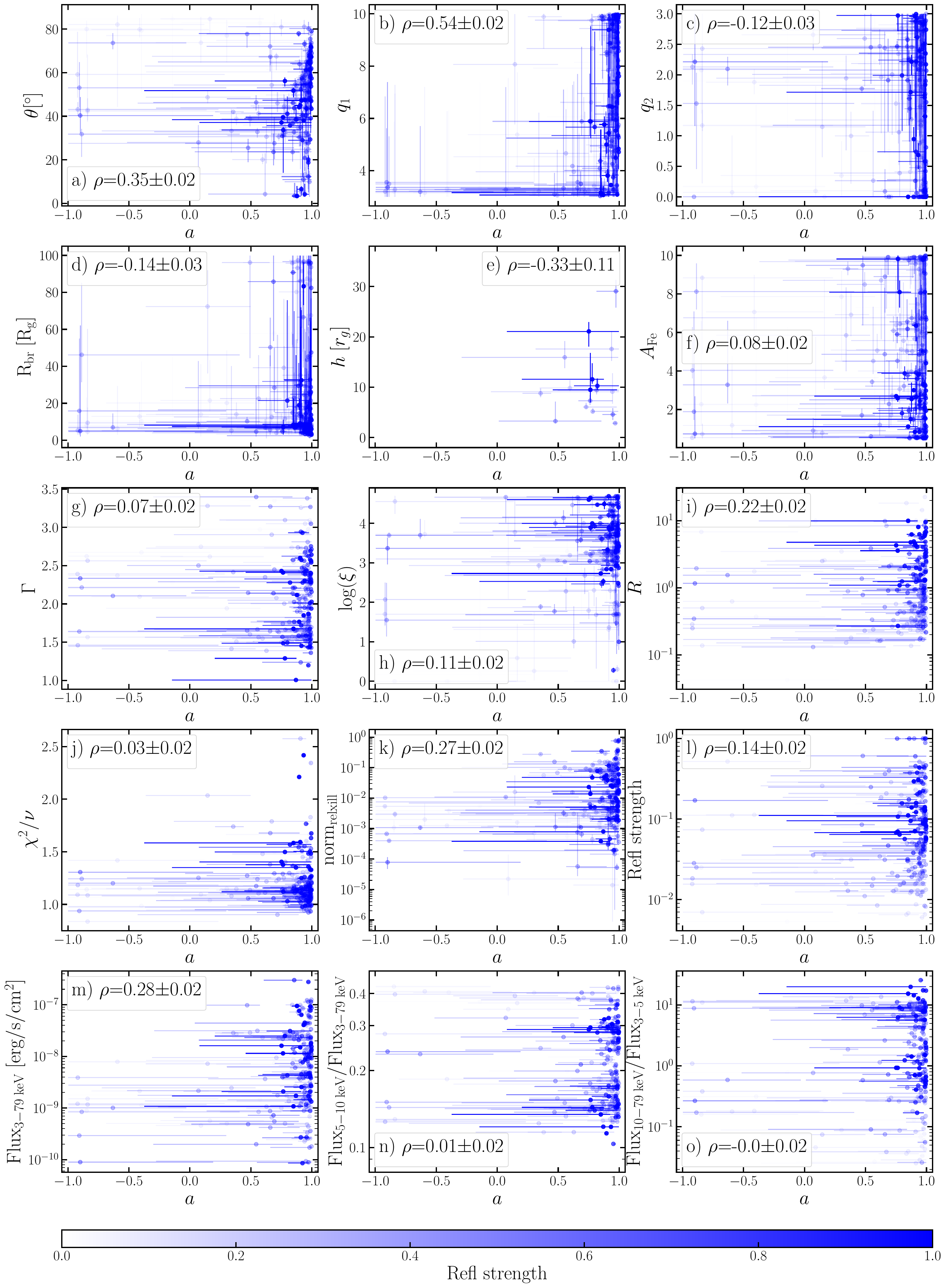}
    \caption{Results of the spectral analysis of the 245 spectra of the 36 sources treated in this work. The x-axis shows the spin measured from each spectrum, while the y-axis represents a few parameters of interest. The transparency of the points represents the strength of the reflection during the observation, defined as the ratio of the reflected flux to total flux in the 3-79 keV band. Note that in the panel showing reflection strength vs. spin, the y values of the points are normalized with respect to the values from other observations of the same source, while the transparency of the points is normalized with respect to values from all observations of all sources. The numbers in the panels represent the Spearman correlation coefficient for the specific combination of parameters.}
    \label{fig:all_vs_spin}
\end{figure*}

We explored some of the correlations of particular interest, both with respect to the BH spin $a$ and between other parameters. The 15 panels in Figure \ref{fig:all_vs_spin} show all the spins measured from all the individual 245 NuSTAR observations analyzed in this work on the x-axis versus a few parameters of interest. The transparency of the points is proportional to the strength of reflection\footnote{The reflection strength is defined as the ratio of the reflected flux to the total flux in the 3--79 keV NuSTAR band.} during the observation, as presented in \cite{2023ApJ...946...19D, Draghis24}. The parameters for which the trends were plotted are the viewing inclination $\theta$, the coronal emissivity parameters $q_1$, $q_2$, and $R_{\rm br}$, the coronal height $h$, the Fe abundance $A_{\rm Fe}$, the power-law index $\Gamma$, the ionization parameter $\log(\xi)$, the reflection fraction parameter in the \texttt{relxill} family of models $R$, the fit statistic $\chi^2/\nu$, the normalization of the \texttt{relxill} component in the fit $\rm norm_{\rm relxill}$, the reflection strength, as defined in \cite{2023ApJ...946...19D}, the 3-79~keV flux, the ratio of the flux in the 5-10~keV band to the flux in the 3-79~keV band, and the hardness ratio of the source during the observation, defined as the ratio of the flux in the 10-79~keV band to the flux in the 3-5~keV band. 

The $\theta-a$ panel in Figure \ref{fig:all_vs_spin} suggests that there is no clear trend or degeneracy between the BH spin and the inclination of the inner accretion disk. This is particularly interesting because both parameters act to distort the shape of the relativistically broadened Fe line. The $q_1-a$ panel shows an interesting trend: spectra favor a low or negative spin only when the inner emissivity index is low, and only for observations where the strength of reflection is low. This translates to the same phenomenon that was discussed in \cite{2023ApJ...954...62D}, where low-quality data have difficulty distinguishing between solutions with high spin and high $q_1$ and low spin and low $q_1$. When $q_1$ is reduced, the emissivity is flatter with radius. In order to maintain the same overall reflected flux, the fits attempt to push the inner disk radius outward. As the inner disk radius is fixed to the ISCO, this translates to an increase in the ISCO, translating to a decrease in spin. These results should be interpreted as a lower limit on spin characterization. In the lamppost geometry, this solution would be analogous to a large coronal height. Similarly, the $q_2-a$ panel shows that generally low spins prefer higher $q_2$, closer to the value of $q_2\sim3$. 

The correlation matrix shown in Figure \ref{fig:correlation_matrix} shows a moderate negative correlation between coronal height and BH spin. However, this correlation is not as extreme as it might initially seem, as it is driven by a single data point with $a\sim-1$ and $h\sim200\;r_g$, obtained in an observation with very low reflection strength. Therefore, we omitted this point from the $h-a$ panel in Figure \ref{fig:all_vs_spin}. The points shown in this panel suggest that, in order to obtain a reliable spin measurement (i.e. low uncertainties) using the lamppost geometry, a low coronal height is required. The lamppost geometry is the preferred choice among the array of models tested throughout this analysis, generally, for cases where the quality of the data is reduced (i.e., low reflection strength). Additionally, it is apparent that the size of the uncertainty on the spin measurement is increasing with increasing coronal height.

Both the flux vs. spin and \texttt{relxill} normalization vs spin panels in Figure \ref{fig:all_vs_spin} convey a powerful insight: low spins are preferred only in observations where the source flux (and therefore the normalization of the \texttt{relxill} component) is low. Observations taken when the sources are bright preferentially produce high spins. Note that for both panels, the y-axis is in logarithmic scale. This point is related to the patterns emerging from the $q_1-a$ panel, where faint observations have trouble distinguishing between families of solutions that produce similar statistics. As the parameter space is wider for the low-spin, low-$q_1$ solution versus the high-spin, high-$q_1$ one, even if the statistic is similar, the walkers in the MCMC analysis will preferentially navigate to the former solution and have difficulty navigating back to the high-spin one. In bright observations, the difference in statistics between the two families of solutions is more significant, forcing walkers to navigate back to the high-spin solutions. Again, these panels emphasize the requirement to obtain multiple observations of sources in order to fully characterize the systematic uncertainties associated with the individual measurements. 

The reflection fraction $R$ versus the spin panel in Figure \ref{fig:all_vs_spin} confirms that high values of $R$ are only favored for high spins. The interpretation of this trend is largely geometric: high spins allow a low coronal height (or a high inner emissivity index $q_1$ when the geometry of the corona is not assumed a priori), leading to more of the emission from the corona being reprocessed by the disk atmosphere. The reflection strength vs. spin panel corroborates the intuition gained from the previous panel. It is important to note that the transparency of the points in all panels shows the strength of reflection during the observations, similar to the y-axis in the reflection strength vs. spin panel. However, counterintuitively, the transparency of the points does not directly increase with increasing y value in the panel. The reasoning behind this is how the values were normalized for plotting vs. for use in the analysis: the y values in the figure represent the reflection strength normalized to the other observations of the same source, ensuring that the sum of the reflection strength of all the observations of the same source equals 1. In contrast, the transparency of the points represents the reflection strength normalized to all values within the sample, so that the highest value in the sample is equal to 1, and the lowest equal to 0. This choice was made for the clarity of the plots and to enable more intuitive comparisons throughout the sample. 

The panel showing the fit statistic ($\chi^2/\nu$, where $\nu$ represents the number of degrees of freedom in the spectral fit) versus measured spin confirms that generally the quality of the fit does not necessarily correlate with the size of the uncertainty in the measurement of the spin. Additionally, most of the low-spin measurements actually come from good spectral fits, with $\chi^2/\nu\sim1$, again confirming that low spins generally emerge from spectra with lower SNR, which are easier to fit. Furthermore, this panel illustrates the ability of our automatic fitting algorithm to identify good-fit solutions. 

Lastly, the lower-right panel in Figure \ref{fig:all_vs_spin} shows the hardness of the source during the respective observations versus the measured BH spin. As our models link the inner disk radius to the ISCO radius, the distribution of points suggests that there is no preferential evidence of disk truncation in the hard states compared to softer spectral states. Such an effect would be apparent through an overabundance of lower spin measurements in certain spectral states. This panel hints at the fact that the accretion disk is extending close to the ISCO for all observations treated in this analysis, regardless of their hardness. The observations analyzed were selected based on the Eddington fraction of the source during the observation, requiring $10^{-3}\leq L/L_{Edd}\leq0.3$. Furthermore, it is noticeable that, as expected, in softer, disk-dominated spectral states, the strength of reflection is generally lower. This is owing to the fact that the disk continuum contributes a larger fraction of the total observed flux during softer states. It is likely that the poor spin constraints are associated with short exposures of lower flux sources, rather than with evidence of disk truncation. Note that all low or negative individual spins have large uncertainties and low reflection strength, meaning that they will not drive the reported combined measurement but they will influence the size of the systematic uncertainty determined.

\subsection{Density Experiments} \label{sec:density}
The $A_{\rm Fe}-a$ panel in Figure \ref{fig:all_vs_spin} suggests that there is no clear trend between the spin of BH and the Fe abundance measured. When looking at the Fe abundance, it becomes apparent that the points are distributed in two general regions of the parameter space, forming two sets of measurements: those that predict low $A_{\rm Fe}$, and those that predict very high $A_{\rm Fe}\sim10$ times that of the Sun. \cite{2018ApJ...855....3T} suggested that for observations of Cygnus X-1, an enhanced accretion disk density of the order of $\rm n=4\times10^{20}\;\rm cm^{-3}$ results in better fits to the data and an Fe abundance close to the solar value, whereas most independent measurements predict an abundance of $A_{\rm Fe}\sim6$. We attempted to investigate this by testing multiple values of the disk density in our models throughout the entire analysis. When attempting to break down the Fe abundance by accretion disk density, no clear trend emerges. However, the upper limit of the parameter space allowed by the models is $n=10^{19}\;\rm cm^{-3}$, which is probably too low to properly account for the elevated Fe abundances seen in many sources. 

Newer versions of the \texttt{relxill} models allow higher accretion disk densities. We fit two observations of GX 339-4, in the soft and hard states, with the \texttt{relxillCp} variant of version 2.3 of \texttt{relxill}, by fixing the accretion disk density to the maximum allowed value in the model, $n=10^{20}\;\rm cm^{-3}$. The quality of the fit was effectively unchanged, the spin and inclination measurements were fully consistent, and the measured Fe abundance was reduced with an increased disk density. This experiment further confirms that the assumption regarding the accretion disk density does not significantly influence the ability to measure the BH spin. 

\begin{figure}[ht!]
    \centering
    \includegraphics[width= 0.45\textwidth]{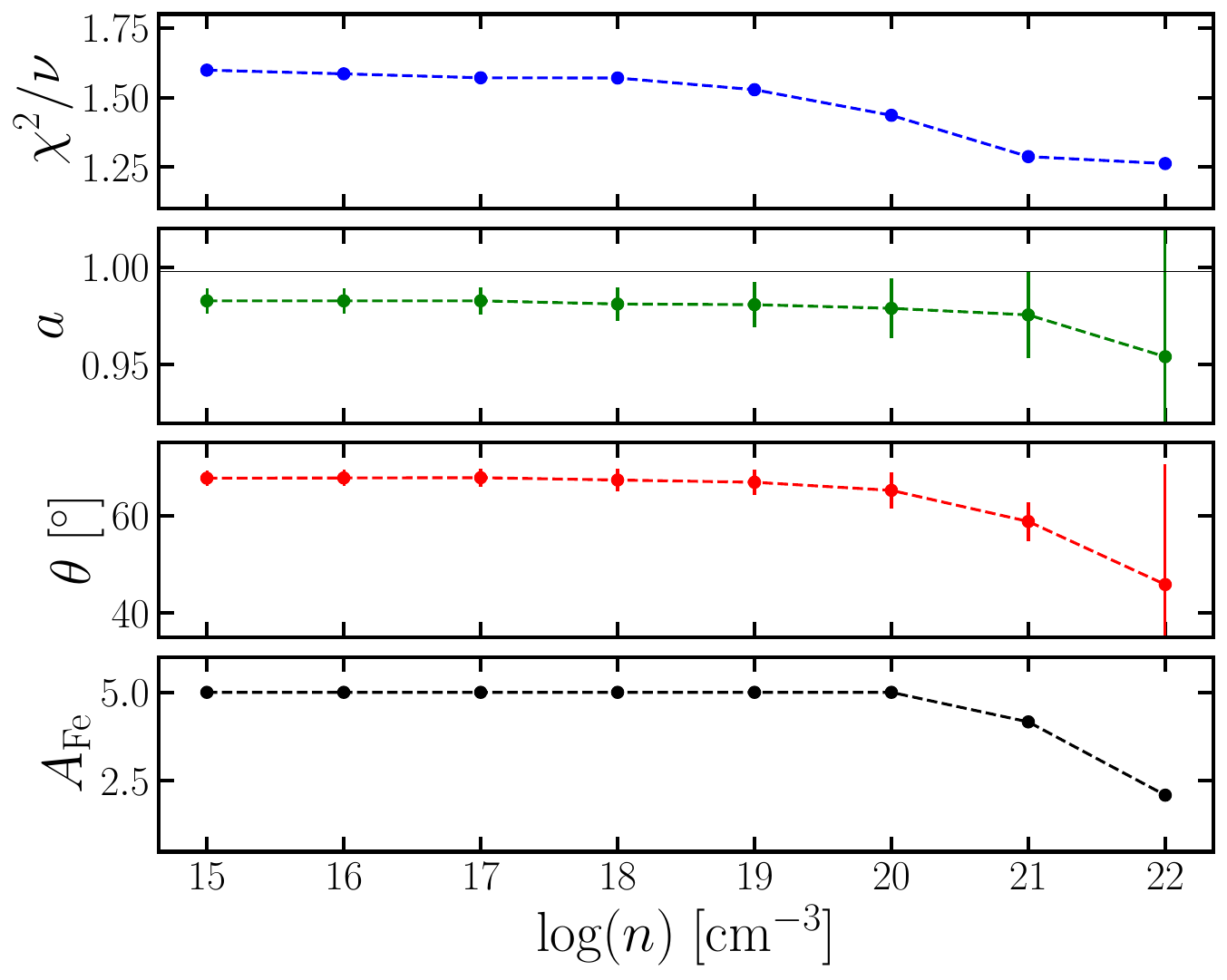}
    \caption{Statistic ($\chi^2/\nu$), BH spin ($a$), viewing inclination ($\theta$), and Fe abundance ($A_{\rm Fe}$) measured by fitting the NuSTAR spectra of MAXI J1820+070 from ObsID 90401309023 with the \texttt{reflionx\_HD} model, with disk density fixed at multiple values across the entire allowed parameter space.}
    \label{fig:pars_vs_logn}
\end{figure}

We further expanded this experiment to one of the very high SNR observations of MAXI J1820+070 in the intermediate state (ObsID 90401309023). The trend remained largely similar to the tests for the GX 339-4 data, with densities up to $10^{20}\;\rm cm^{-3}$ not producing significantly different results. We further pushed this experiment by testing the impact of using the reflection model implemented and tested in \cite{2018ApJ...855....3T}: \texttt{reflionx\_HD}. This model allows for disk densities up to $10^{22}\;\rm cm^{-3}$. We fit the data with the density fixed at integer increments throughout the allowed parameter space of $\log(n)=15-22$. The best-fit $\chi^2/\nu$, measured spin, inclination, and Fe abundance for fits with the varying densities are shown in Figure \ref{fig:pars_vs_logn}. The uncertainties in $A_{\rm Fe}$ are not shown in Figure \ref{fig:pars_vs_logn} as they cannot be reliably constrained from a simple fit without properly assessing the complete shape of the parameter space through an MCMC run, and the uncertainties in $a$ and $\theta$ are calculated as the square root of the diagonal entry in the covariance matrix of the fit, representing the values produced by simple \texttt{xspec} fits.  It is immediately noticeable from the top panel that higher densities produce better fits, but the improvement becomes more significant for $\log(n/\rm cm^{-3})\gtrsim20$, where the spin, inclination, and Fe abundance begin to become more affected. The best statistic is obtained for $\log(n/\rm cm^{-3})=22$, the upper bound of the parameter in the model. For this density, which is 7 orders of magnitude larger than the default value in \texttt{relxill}, the Fe abundance is significantly lower. The high-density model predicts slightly lower spin measurements, however, with larger uncertainties that are fully consistent with the measurements obtained using any disk density value. In addition, these values are also within the magnitude of the systematic uncertainties associated with combining measurements from multiple data sets. Increasing the disk density also leads to lower measured inclinations. This is likely due to the connection between the disk density, the Fe abundance, and the ionization parameter within the models, which impact the inclination constraint, as they all act on the blue wing of the Fe line. Ultimately, we conclude that the assumption of disk density, regardless of the model adopted, does not significantly impact the spin measurement but can influence the inferred inclination. 
\newline
\newline

\subsection{Other Correlations} \label{sec:other_corr}

\begin{figure*}[ht!]
    \centering
    \includegraphics[width= 0.9\textwidth]{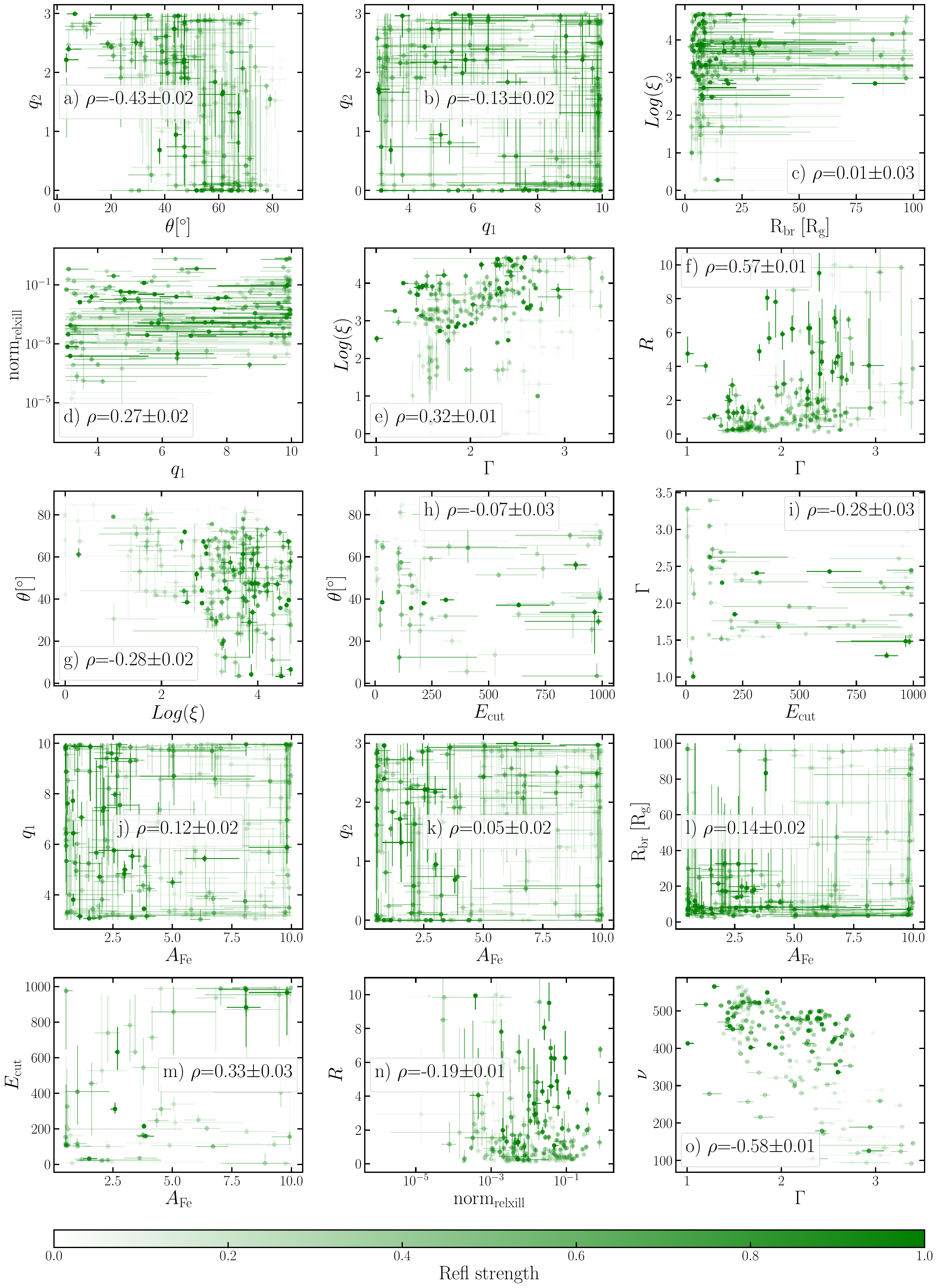}
    \caption{Interesting parameter combinations emerging from the analysis of the 245 spectra treated in \cite{Draghis24}. The transparency of the points is proportional to the strength of reflection during each of the observations that produce the individual points. The trends emerging from this figure highlight the complexity of the parameter space and potentially hold the key to completely understanding the behavior of the models used in measuring BH spin using relativistic reflection. The numbers in the panels represent the Spearman correlation coefficient for the specific combination of parameters.}
    \label{fig:interesting_plots}
\end{figure*}

In addition to characterizing the global impact of the other parameters in the fit on the ability to constrain the BH spin, this data set allows us to quantify the ability to constrain other parameters of interest and to probe for potential trends regarding the physical behavior of the systems. Figure \ref{fig:interesting_plots} shows a few combinations of parameters that show particularly interesting trends. There are strong negative global correlations between the outer emissivity index $q_2$ and the inclination of the inner accretion disk $\theta$, and between the inclination and the ionization parameter $\log(\xi)$. At the same time, the ionization parameter is positively correlated with the power-law photon index $\Gamma$ and with the breaking radius of the emissivity profile $R_{\rm br}$. Other positive correlations include the one between the normalization of the \texttt{relxill} component and the inner emissivity index ($q_1$), again linking to the difficulty of distinguishing between high-spin, high-$q_1$ and low-spin, low-$q_1$ solutions in spectra of faint sources. Furthermore, a positive correlation is apparent between the power-law photon index $\Gamma$ and the reflection fraction $R$, but the large uncertainties at large values for both parameters make it difficult to assess whether this is a physical trend. A clear trend is seen between the number of degrees of freedom in the spectral fit $\nu$ and the power-law photon index $\Gamma$, which can be explained by the fact that harder spectra generally become dominated by the background at lower energies, leading to fewer spectral bins at high energies and, therefore, lower $\nu$.

An interesting trend appears in the $E_{\rm cut}-A_{\rm Fe}$ panel in Figure \ref{fig:interesting_plots}: higher Fe abundances are measured when the cutoff energy of the incident power law is higher. This connects to the slight negative correlation observed in the $\Gamma-E_{\rm cut}$ panel, which in turn relates to the strong positive correlation seen between the ionization parameter and the power-law photon index $\log(\xi)-\Gamma$. However, looking at $A_{\rm Fe}$ versus $\log(\xi)$ shows no clear trend, highlighting the complexity of the parameter space. Understanding how all the parameters relate to each other is a challenging process, and this data set holds the key to fully unlocking the behavior of the models. In the future, implementing the results of high-resolution spectroscopy and X-ray polarization measurements into the information provided by this dataset will facilitate breaking degeneracies between parameters by providing reliable, independent measurements of certain parameters such as the inclination, ionization, and geometry of the corona.

\newpage

\subsection{Trends in Individual Sources}\label{sec:individual}

\begin{figure}[ht]
    \centering
    \includegraphics[width= 0.45\textwidth]{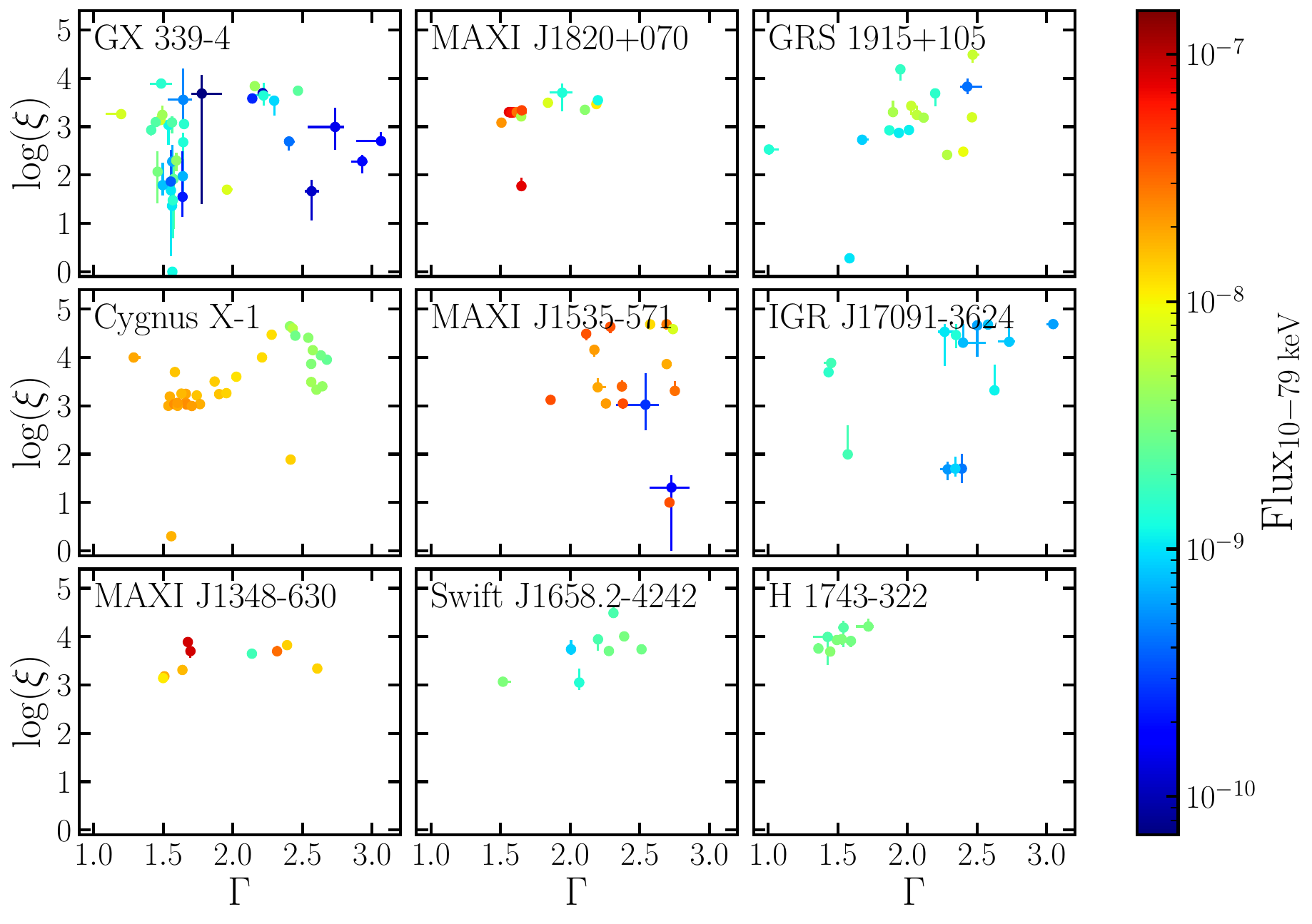}
    \caption{The nine panels show the $\log(\xi)$ and $\Gamma$ values measured in each individual observation in our analysis for nine sources that have multiple observations. The colors of the points represent the 10--79~keV flux during each individual observation.}
    \label{fig:xi-vs-gamma-flux}
\end{figure}

While it is appealing to treat all spectra in a source-independent way, it is important to acknowledge that each of the 36 systems treated in this analysis is unique in its own regards. We tried to characterize whether the apparent trend in the $\log(\xi)-\Gamma$ panel in Figure \ref{fig:interesting_plots} is also noticeable in observations of individual sources. Figure \ref{fig:xi-vs-gamma-flux} shows the same parameter combination for nine sources with multiple observations: GX 339-4, MAXI J1820+070, GRS 1915+105, Cygnus X-1, MAXI J1535-571, IGR J17091-3624, MAXI J1348-630, Swift J1658.2-4242, and H 1743-322. The points in Figure \ref{fig:xi-vs-gamma-flux} are colored to represent the flux in the 10--79 keV band, as the high-energy X-rays produce the ionization of Fe. With the exception of Cygnus X-1, none of the sources show clear trends of increasing ionization with increasing power-law index, or any clear dependence on ionizing flux. The observations of Cygnus X-1 present the aforementioned correlation between the parameters, with increasing ionization in softer states. Moreover, a lower $\Gamma$ corresponds to a higher 10--79 keV flux, expected in harder states. However, the trend between ionization and flux is counterintuitive: observations during which the ionizing flux is higher lead to lower values of the ionization parameter. Softer spectral states tend to produce spectra in which the measured disk ionization is higher, consistent with the presence of a hotter accretion disk. Although Cygnus X-1 is a unique source in many respects, this trend illustrates the complexity of the parameter space and is likely connected to the assumption of a low and constant density of the accretion disk atmosphere. However, this figure illustrates that the trend that emerges in the $\log(\xi)-\Gamma$ panel in Figure \ref{fig:interesting_plots} is not uniquely influenced by observations of one source and rather demonstrates the overall behavior of the models. 

\begin{figure}[ht]
    \centering
    \includegraphics[width= 0.45\textwidth]{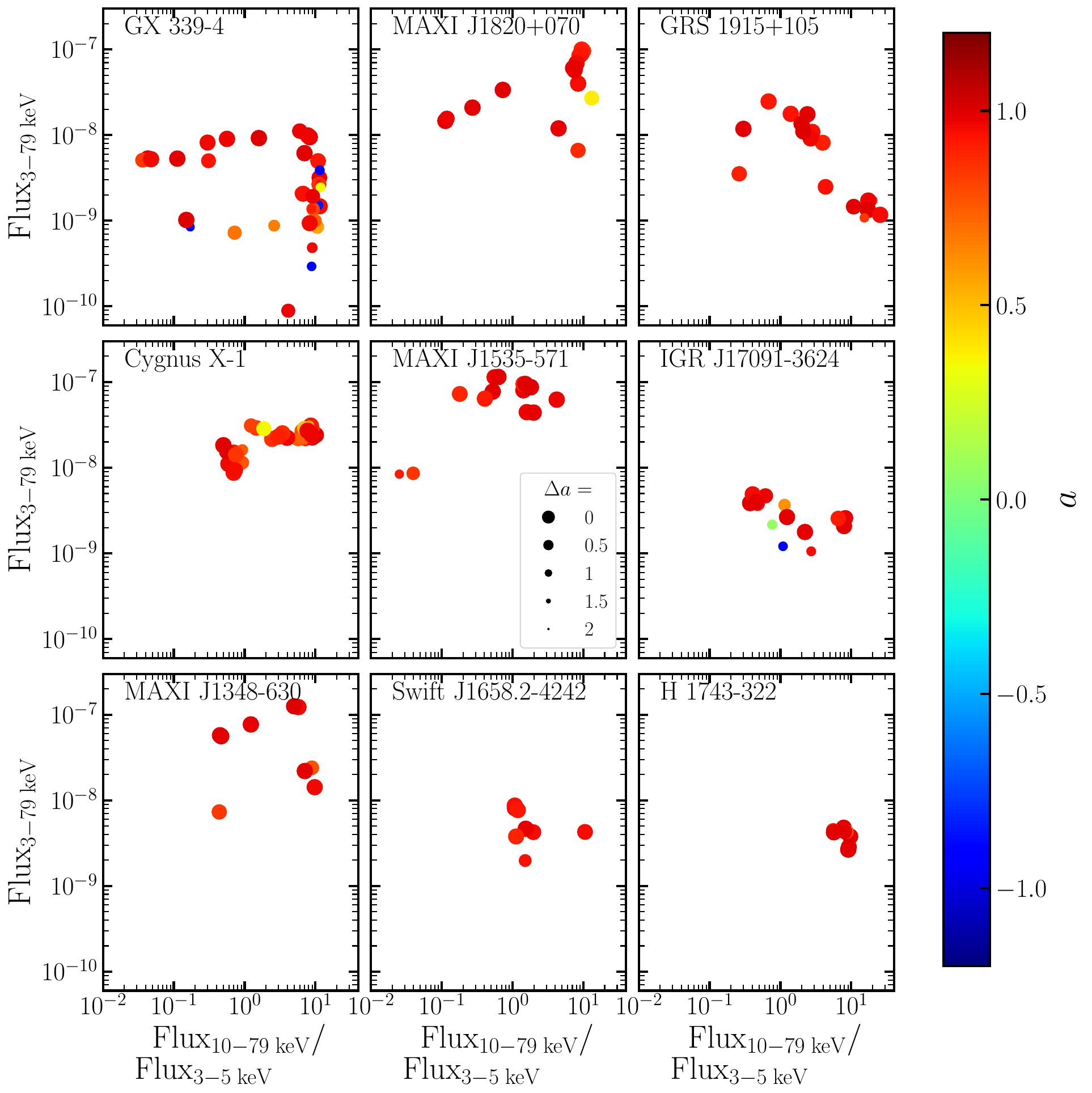}
    \caption{Hardness-intensity diagrams for nine sources with multiple observations. The color of the points represents the measured value of the BH spin, and the size of the points represents the uncertainty in the spin measurement, as shown in the central panel, with larger markers representing smaller measurement uncertainties.}
    \label{fig:HIDs}
\end{figure}

Lastly, while the goal of this analysis was to provide a source-independent view of the spectral fits, there is undoubtedly value in looking at individual sources. Figure \ref{fig:HIDs} shows hardness-intensity diagrams for the same nine sources with multiple observations. The hardness in this figure is defined as the ratio of the 10--79~keV flux to the 3--5~keV flux, while the intensity is represented by the flux measured in the entire NuSTAR 3--79~keV bandpass. The color of the points indicates the best-fit measured spin, and the size of the points is inversely proportional to the uncertainty of the spin measurement. Given that for each source the spin has a singular value, this figure is informative with regard to our ability to constrain it as a function of the spectral state in which the source was during the observation analyzed and of the source flux. Generally, the measurements with low uncertainties agree on high spins, while the low spins (bluer points) are generally associated with worse constraints (smaller markers). This is particularly noticeable in the top-left panel, representing the measurements for GX 339-4. It is interesting to note that low spins with high uncertainties (small, blue points) are measured from spectra obtained under conditions very similar to those in conditions that produce tightly constrained, high spins (larger, red points). This suggests that the flux or the hardness are not the main drivers of the ability to constrain spin. 

\subsection{Effects of Flux and Hardness}\label{sec:flux_and_hardness}

\begin{figure}[ht]
    \centering
    \includegraphics[width= 0.45\textwidth]{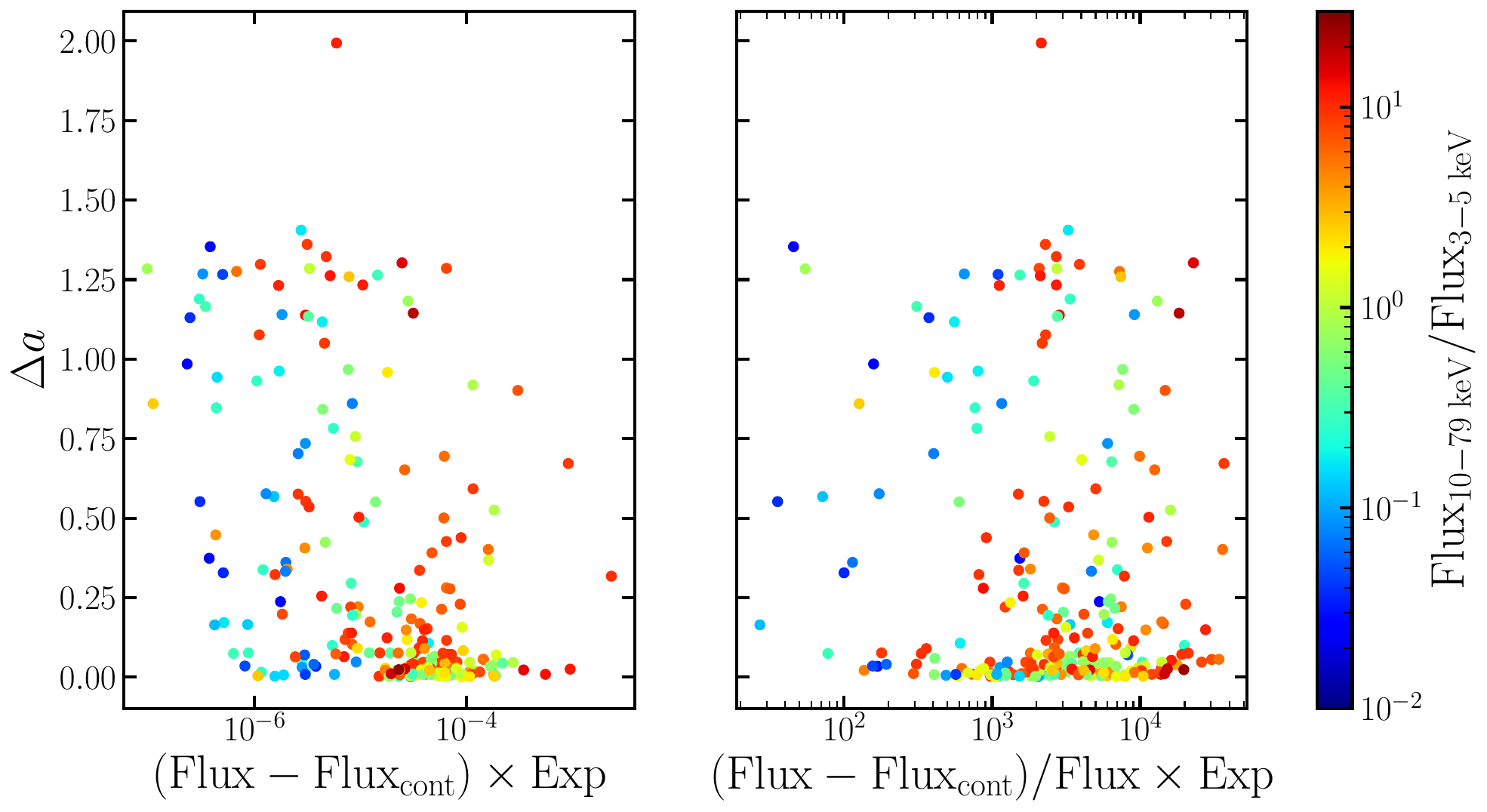}
    \caption{Left: $1\;\sigma$ uncertainty in spin measurement in all observations in this study vs. the number of counts in the reflection component, calculated as the difference between the total flux during the observation and the flux in the underlying continuum, multiplied by the exposure of the observation. Formally, the units of the quantity on the x-axis are of $\rm erg/cm^2$. The color of the points represents the hardness ratio of the source, with redder points representing observations taken while the sources were in harder states, whereas bluer points represent softer spectral states. Right: this panel shows the values in the left panel, divided by the flux of the source during the corresponding observation plotted. Formally, the units of the quantity on the x-axis are seconds. However, the quantity represents the ability to disentangle the reflected spectrum from the underlying continuum emission.}
    \label{fig:da_vs_counts}
\end{figure}

This trend is further explored in Figure \ref{fig:da_vs_counts}, where we plot the uncertainty of the spin constraint versus a proxy for the number of counts within the reflected spectrum. The left panel shows the difference between the total 3--79~keV flux and the ``continuum" flux in the same band, calculated by setting the reflection fraction within the models to zero. This difference corresponds to the amount of flux owing only to the reflected component and is then multiplied by the exposure of the observation. In the right panel, we divide the quantities in the left panel by the 3--79~keV flux, in order to illustrate the ability to disentangle the reflected spectrum from the underlying continuum emission. The colors of the points represent the hardness of the spectrum, calculated as the ratio of the 10--79~keV flux to the 3--5~keV flux. The left panel suggests that the uncertainty in the spin measurement increases as the total number of counts in the reflected spectrum decreases, and is also correlated with softer spectra (bluer points). The takeaway of the right panel is not as straightforward, and it suggests that the uncertainty in spin is not immediately correlated with the fraction of the overall total number of counts that is attributed to reflection, meaning that the limiting factor in increasing the precision of spin measurements is not the ability to disentangle the reflected flux from the underlying continuum emission but rather the total number of counts that we can detect that are associated with the reflected emission.  
\newline
\newline

\subsection{Impact of Absorption Lines}\label{sec:lines}
In the fitting of the 245 NuSTAR spectra, when statistically motivated, narrow Gaussian additive components with negative normalizations are included in the model around $\sim7\;\rm keV$. These components aim to account for absorption features in the spectra caused by ionized winds, likely launched from the accretion disk. For Fe XXV (with a rest energy of $\sim6.7$ keV) and Fe XXVI (with a rest energy of $\sim6.9$ keV), these absorption features land on the blue wing of the relativistically broadened Fe K line. Such features have previously been observed in high-resolution spectra of accreting BH XBs such as GRO J1655-40 (\citealt{2008ApJ...680.1359M}) and GRS 1915+105 (\citealt{2016ApJ...821L...9M}). The presence or absence of those features, and more importantly our ability to properly quantify their magnitude in the moderate-resolution NuSTAR spectra, has the potential to significantly influence a number of parameters in the spectral fits, including the BH spin, inclination, ionization parameter, etc. 

In our analysis, the inclusion of absorption features in the models is motivated by the data and by the quality of the spectral fits. If the improvement in fit statistic is significant, the absorption features are included in the models for the individual spectrum. This implies that if the models do not include a negative Gaussian component, the spectra do not show evidence of the presence of absorption features. However, it is commonly acknowledged that the lack of evidence does not represent evidence of absence. In this case, the fit statistic that does not improve significantly following the addition of absorption components can simply suggest that the SNR of the spectra cannot reliably quantify the presence of such absorption features. Therefore, there is a possibility that such absorption features are indeed present in the spectra but not properly accounted for because of the limited SNR of the data. 

\begin{figure}[ht!]
    \centering
    \includegraphics[width= 0.45\textwidth]{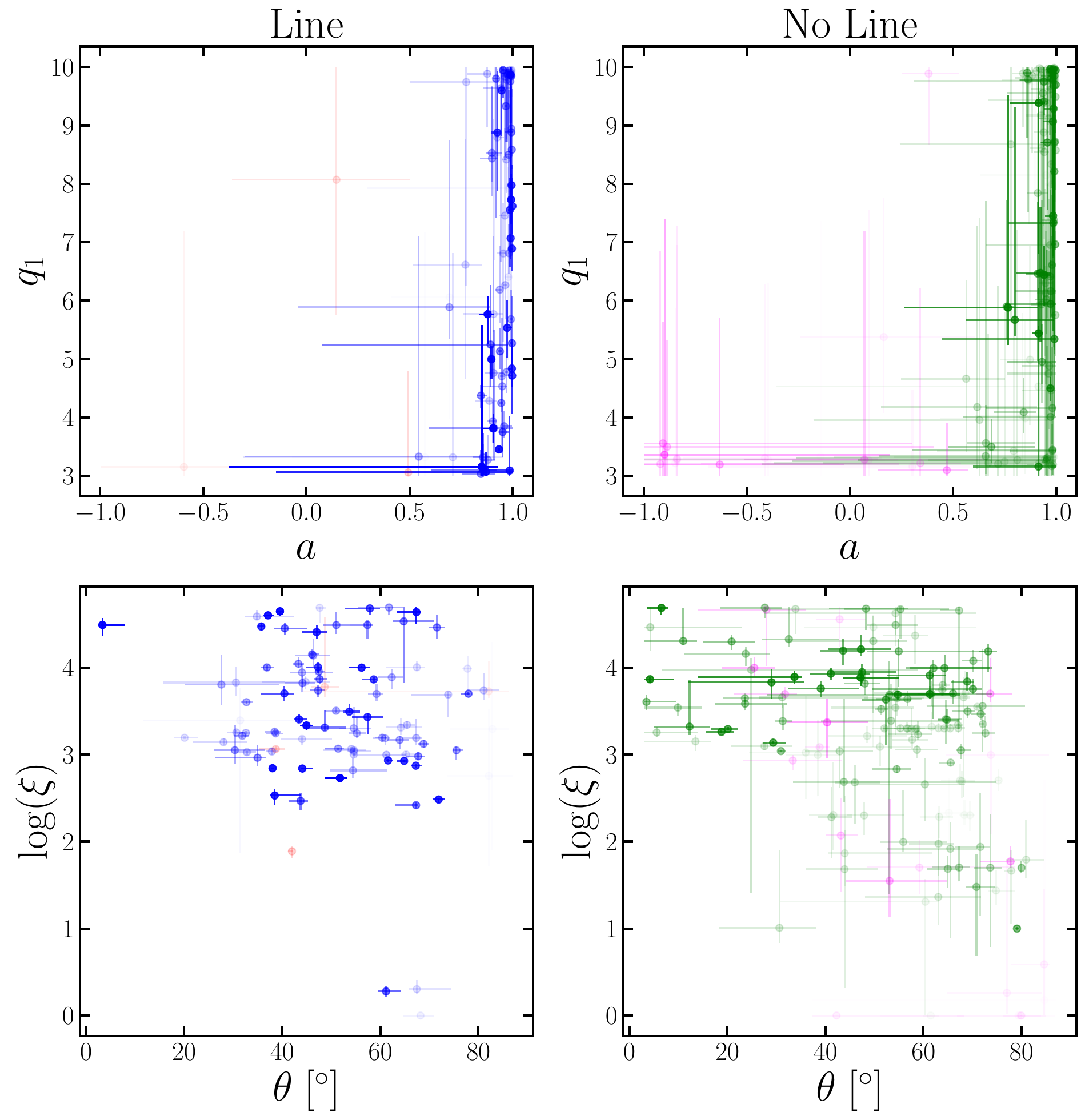}
    \caption{The measured inner emissivity index $q_1$ vs. BH spin $a$ (top) and ionization parameter $\log(\xi)$ vs. inclination of the inner accretion disk $\theta$ (bottom) produced from the models that include an absorption \texttt{gaussian} component around 7 keV (left) and that do not include an absorption component (right). Note that the left and right panels show results from different spectra, using the best-performing models for the given spectrum. The inclusion of the absorption features is dictated by a significant improvement in the fit statistic. The red points (left) and magenta points (right) in both the upper and lower panels show results from spectra for which the measured spin is $a<0.5$. Similarly to Figures \ref{fig:all_vs_spin} and \ref{fig:interesting_plots}, the transparency of the points is proportional to the strength of reflection during each individual observation.}
    \label{fig:Line_presence}
\end{figure}

In an attempt to quantify the effects of the inclusion of negative Gaussian components in our models, we split the results of the spectral fitting of the 245 NuSTAR observations based on the presence or absence of the Gaussian component in the model. The clearest way to illustrate the impact of this division is shown in Figure \ref{fig:Line_presence}. The left panels of Figure \ref{fig:Line_presence} show the parameters obtained from fits where a Gaussian line was included in the models, while the right panels show the parameters obtained from fits where absorption lines were not included in the models. Note that each individual point represents the result from the analysis of a different spectrum and that this figure does not compare the results obtained by fitting the same data sets with different models. In our analysis, 96 spectra required an absorption line (left in Figure \ref{fig:Line_presence}) and 149 did not show a significant improvement in the fit statistic by including an absorption line (right in Figure \ref{fig:Line_presence}), either because there was no absorption in the system or because the quality of the data did not allow significant constraints of those features. The upper panels show the $q_1-a$ parameter space, while the lower panels show the $\log(\xi)-\theta$ parameter space. In the left panels, the blue points represent observations for which the mode of the measured posterior distribution for the BH spin was $\geq0.5$, and the red points represent observations for which $a<0.5$ was measured. Similarly, in the right panels, the green points show $a\geq0.5$, while the magenta points show $a<0.5$. 

When an absorption line is present in the models, the fits almost never favor low or negative spins, and most of the low-spin measurements are produced by models that do not contain an absorption line (the magenta points in the top-right panel in Figure \ref{fig:Line_presence}). At the same time, the low spins in the ``no line" case generally also have low inner coronal emissivity $q_1$. This combination of low spin, low $q_1$ has been discussed in Section \ref{sec:spin_corr}, and probably illustrates that when the quality of the data is low, the presence of lines cannot be assessed, and also the fits have trouble distinguishing between low-$q_1$, low-$a$ solutions and high-$q_1$, high-$a$ solutions. This is likely not a correlation, but rather just a consequence of using low-SNR spectra. Still, when the quality of the data is high enough to enable a reliable assessment of the presence of an absorption line, the measured spin is almost never low. 

The lower panels in Figure \ref{fig:Line_presence} highlight another interesting trend. Low ionizations $\log(\xi)\lesssim2.5$ are predominantly measured when the models do not include an absorption line, with almost no instances of low ionization being predicted by models that include absorption features. At the same time, low ionizations tend to also produce higher inclinations $\theta\gtrsim45^\circ$ (when no line is included in the model), while in the case when the model includes a Gaussian component, no small inclinations $\theta\leq20^\circ$ are measured with the exception of the fit to the MAXI J1535-571 spectrum from ObsID 80302309006. It is likely that when an absorption feature is improperly accounted for, the ionization parameter and the inclination account for the sharp drop in the blue wing of the relativistically broadened Fe line by artificially reducing the ionization and increasing the inclination. At the same time, this trend could actually have physical origins, with disks in which the atmosphere is less ionized being less likely to launch winds that would produce absorption features. In this case, the inclinations of those low-ionization sources are simply distributed around the median orientation angle for isotropically distributed orbits of $60^\circ$, with a slight preference for higher inclinations, as those tend to produce narrower Fe K lines, and therefore concentrate the flux in fewer spectral bins and making a reflection detection more likely. Furthermore, it is possible that due to geometric arguments, we do not see winds in sources with low inclination, explaining the presence of the points at low $\theta$ in the lower-right panel of Figure \ref{fig:Line_presence}.

It is interesting to note that the observations that produce the low-spin, low-$q_1$ points in the ``no line" case (pink points in the top-right panel in Figure \ref{fig:Line_presence}) are not the same points that cause the discrepancy between the two panels in the bottom row. This suggests that the two trends highlighted above are not related. There are two main takeaways from this experiment. First, if the SNR of the spectra is not high enough to reliably constrain the presence of absorption features, then it is also possibly not high enough to reliably disentangle the low-spin, low-$q_1$ solutions from the high-spin, high-$q_1$ solutions. Second, if the absorption features are not properly accounted for, the inclination and ionization measurements can be affected, which further leads to improper characterization of the disk density, Fe abundance, reflection fraction, etc., as highlighted in Section \ref{sec:other_corr}. This experiment emphasizes the importance of understanding the properties of disk winds in XB systems, studies for which the high-resolution microcalorimeter Resolve on board XRISM (\citealt{2020SPIE11444E..22T}) is ideally suited. 

\section{Summary and Conclusions} \label{sec:summary}

In \cite{Draghis24}, we presented the results of the spectral analysis of 245 NuSTAR observations of 36 BH XB systems, focusing on the measured spin and inclination. In this paper, we further present the results of the spectral analysis of the same data set, focusing on the rest of the parameters and mainly on the potential influence they have on the ability to reliably measure the BH spin. 

In Section \ref{sec:pop}, we show the spins of the 36 BHs in relation to the independently measured system parameters for those sources. 
BH systems shown to undergo outbursts tend to have low masses, generally $\leq15M_\odot$, as previously noted by works such as \cite{2016A&A...587A..61C} and \cite{2021ApJ...921..131J}. The uncertainties in spin measurements appear to decrease with increasing BH mass and increase with distance.
Looking at the orbital periods of the systems and comparing them to the theoretical predictions of \cite{2015ApJ...800...17F} for the maximum spin that a BH can achieve through only accretion, it becomes apparent that a number of BHs, especially at low orbital periods, have spins higher than accretion alone can explain, indicating that the BHs must have already formed rapidly rotating, as predicted by works such as \cite{1999MNRAS.305..654K} and \cite{2023arXiv230409350H}. This has implications for the way we understand the formation of BHs and of XB systems, and for the way these systems evolve, highlighting the importance of a lack of low spins being measured in our sample. 

In Section \ref{sec:distributions} we compare the observed distribution of BH spins generated from the measurements in this sample with the distribution of BH spins inferred based on the third edition of the GWTC. For the first time, this comparison is made based on X-ray measurements performed in a systematic, uniform way, not influenced by differences in assumptions going into different measurements. This result confirms and strengthens the claims of works such as \cite{2022ApJ...929L..26F} and \cite{2023ApJ...946...19D}, that the distribution of spins observed in XBs is incompatible with the distribution inferred based on GW measurements. 

The work presented in Section \ref{sec:pars} delves into the intricacies of parameter correlations in spectral fits. This study is conducted in a source-independent way, where each individual spectrum from the 245 observations of the 36 BH systems is treated as a single data point. The aim of this experiment is to assess the behavior of the spectral models in a way that bypasses the source particularities. In Section \ref{sec:spin_corr} we discuss the impact of other parameters on the measured BH spin, and we find that the main aspect to be aware of is that in low-SNR spectra, the fits may have trouble distinguishing between a narrow region of the parameter space with high $a$, high $q_1$ and a broad region of the parameter space with low $a$, low $q_1$. This is a known issue, and it was articulated previously in \cite{2023ApJ...954...62D}. Most importantly, other parameters do not seem to directly and strongly influence the measured spin. Still, a question remains open regarding the impact of the density of the accretion disk on the measured spin. We tackled this issue in Section \ref{sec:density}, and found that the effects of the disk density tend to be noticeable above $\log(n/\rm cm^{-3})\gtrsim20$, however, the impact on the spin measurement is not significant especially when compared to the systematic uncertainty introduced by simultaneously considering results from multiple observations. 

In Section \ref{sec:other_corr} we briefly present other possible degeneracies between the parameters. We note that some trends between $\log(\xi)-\Gamma$, $\theta-\log(\xi)$, $R-\Gamma$, and $q_2-\theta$ are apparent and potentially crucial to understand, especially when spectra show evidence of absorption features caused by the presence of disk winds in the systems. This phenomenon was explored in Section \ref{sec:lines}, where we examine the fits that prefer the addition of negative Gaussian features to the model versus those that do not. We find that in spectra where the SNR is low, it is possible to simultaneously overlook the presence of absorption features and to encounter the confusion between high-$a$, high-$q_1$ and low-$a$, low-$q_1$ solutions. It is unclear whether the two phenomena are related. The second trend that emerged from this study was that only observations that do not require absorption features allow low-ionization measurements. This can either be a modeling artifact, where the fit attempts to alter the value of the ionization to fully account for the shape of the Fe K line, or it can have a physical explanation that disks with low ionization do not launch winds. Regardless of the reason, given the correlations highlighted here, we emphasize the importance of properly accounting for narrow features in the spectra in an attempt to reliably quantify system properties.

The analysis presented in this paper represents a complex, yet not exhaustive exploration of the parameter space of these spectral fits. By systematically fitting the spectra from 245 NuSTAR observations, we compiled a unique dataset of unprecedented breadth, which has the potential to inform future exploratory studies of accreting BH XBs and to guide the development of future X-ray instruments. To maximize the impact of this dataset, we make the table of results from the spectral fitting available to the public (see link in Section \ref{sec:data}), and encourage future exploration using the results of this work. 

The key remaining challenge in measuring X-ray detected spins is the lack of precise understanding of the disk and corona geometries, which is crucial to distinguishing their features from the underlying continuum X-ray emission and accurately measuring BH spin. For BH spin measurements in XBs, the theoretical models used to explain the observed X-ray spectra rely on simplifying assumptions. These models often overlook narrow spectral features and focus on broader signatures. Furthermore, recent X-ray polarization studies predict coronal geometries that conflict with the simplifying assumptions used in spectral analysis (see, e.g., \citealt{2022Sci...378..650K}).

It is important to note that NuSTAR has a limited bandpass and energy resolution, which could impact the robustness of the measurements. \cite{2023ApJ...954...62D} show that performing reflection studies on joint NuSTAR and NICER observations versus on NuSTAR data alone can improve the spin constraint, especially when the SNR of the spectra is low or the reflection features are weak. However, it is unclear what the impact of narrow, unresolved emission and absorption features is on reflection measurements. Although the broad features of reflection are not directly influenced by narrow features, the ability to constrain the underlying continuum may be impacted, translating to changes in the inferred reflection parameters. Spectra from the Resolve microcalorimeter on board XRISM are guaranteed to clarify this possible limitation of reflection measurements and to quantify the impact on our ability to constrain system properties.

The distribution of BHs observed to merge through GWs is rapidly expanding. The fourth observing run of the LIGO-Virgo-KAGRA collaboration is currently ongoing and has already detected more than 200 BBH merger candidates so far, with the fifth observing run planned to begin around 2027. The launch of the Laser Interferometer Space Antenna (LISA; \citealt{2017arXiv170200786A}) is planned around 2035, together with next-generation ground-based GW detectors such as the Cosmic Explorer (CE; \citealt{2019BAAS...51g..35R}) and the Einstein Telescope (ET; \citealt{2020JCAP...03..050M}). Furthermore, next-generation high-resolution X-ray spectroscopy missions such as NewAthena (\citealt{2025NatAs...9...36C}) will expand the work of XRISM and further advance our understanding of the physics of accretion onto compact objects. Gamma-ray missions such as COSI (\citealt{2024icrc.confE.745T}) will further broaden the horizons of what high-energy astrophysics of compact objects looks like. All these efforts will be aided by the sky monitoring of the Rubin Observatory (\citealt{2019ApJ...873..111I}).

In conclusion, while the analysis presented in this work takes important steps in advancing the study of BH rotation, the future on both long and short timescales holds tremendous promise. Addressing the current limitations in understanding disk and corona geometries is essential to accurately discern features in X-ray spectra and precisely measure BH spins. Integrating high-resolution spectroscopy with timing and X-ray polarization data, facilitated by the IXPE polarimetry mission and the XRISM microcalorimeter spectrometer, promises to refine our understanding of BH properties. As the number of merging BHs detected through GWs increases over the following observing cycles, and as current- and future-generation space- and ground-based electromagnetic observatories continue to expand the sample of known BH systems, the constraints on the characteristics of the BH population will continue to improve. By leveraging the advancements provided by a more detailed characterization of a continuously increasing sample of BHs, future research endeavors will continue to unravel the mysteries surrounding BHs and enhance our understanding of the most enigmatic phenomena in the Universe.

\begin{acknowledgments}
The authors thank the anonymous reviewer for their feedback and suggestions, which have improved the structure and contents of this manuscript. P.D. acknowledges helpful conversations with Erin Kara, Kevin Burdge, and Matthew Mould. A.Z. is supported by NASA under award number 80GSFC24M0006.

\textit{Software:} 
\texttt{numpy} (\citealt{harris2020array}), 
\texttt{matplotlib} (\citealt{Hunter:2007}), 
\texttt{scipy} (\citealt{2020SciPy-NMeth}), 
\texttt{Astropy} (\citealt{astropy:2013, astropy:2018, astropy:2022}), 
\texttt{pandas} (\citealt{reback2020pandas, mckinney-proc-scipy-2010}), 
\texttt{corner} (\citealt{corner}), 
\texttt{emcee} (\citealt{2013PASP..125..306F}), 
\texttt{Xspec} (\citealt{1996ASPC..101...17A}), 
\texttt{relxill} (\citealt{2014MNRAS.444L.100D, 2014ApJ...782...76G}).
\end{acknowledgments}


\newpage
\appendix
\section{Complete Correlation Matrix} \label{sec:full_corr_matrix}

\begin{figure}[h]
    \centering
    \includegraphics[width= 0.9\textwidth]{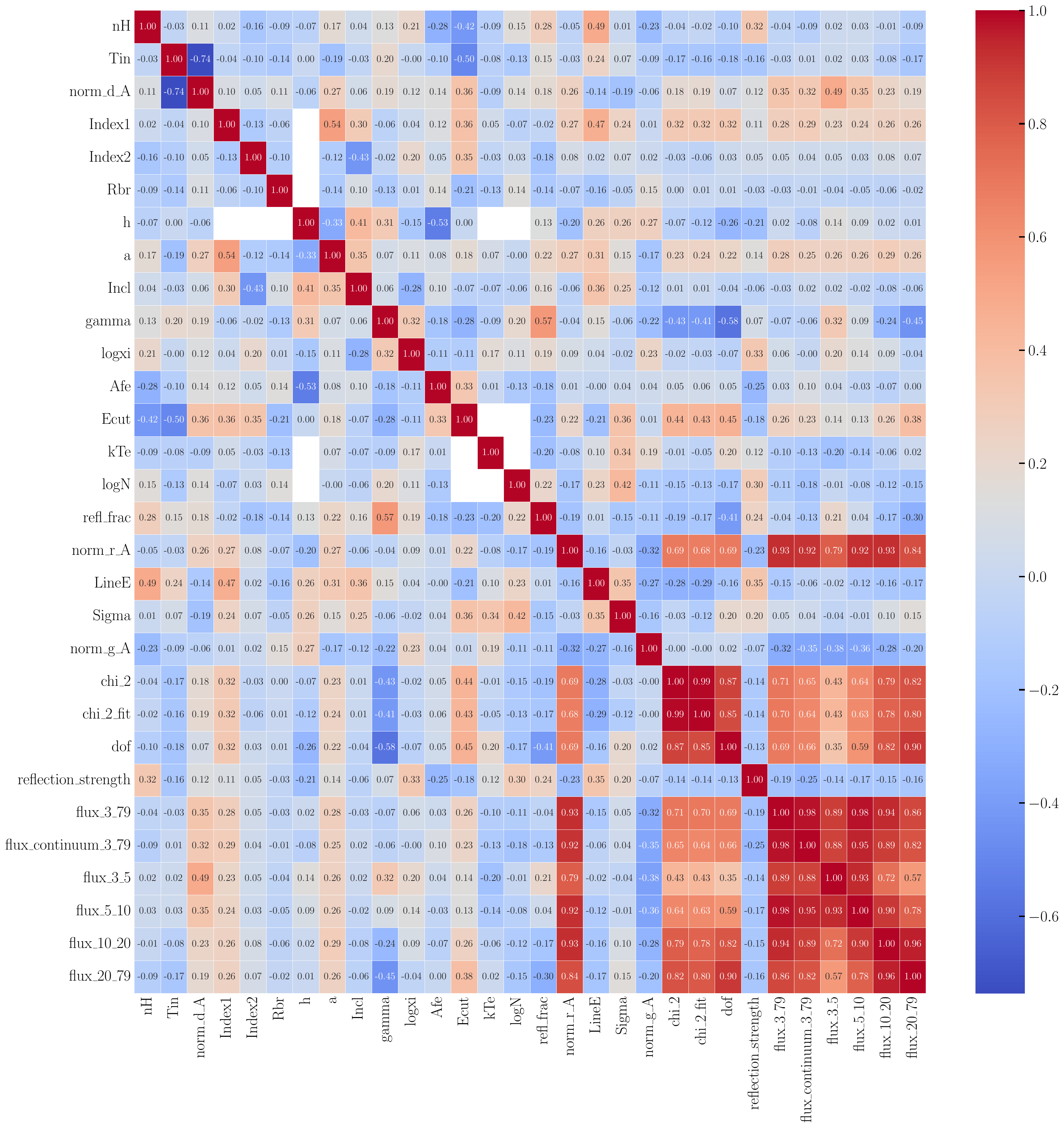}
    \caption{Complete correlation matrix of all the parameters in the spectral models used to fit the 245 NuSTAR spectra presented in this work. The numbers in each cell represent the Spearman correlation coefficient for the given combination of parameters. Redder colors represent stronger positive correlations and bluer colors indicate stronger negative correlations. }
    \label{fig:corr_matrix_complete}
\end{figure}

In section \ref{sec:pars} we show a correlation matrix of the parameters in the \texttt{TBabs}, \texttt{diskbb}, and \texttt{relxill} components in the spectral models used to fit the 245 NuSTAR observations of the 36 BH XB systems treated in this work. Here (Figure \ref{fig:corr_matrix_complete}), we show a more comprehensive correlation matrix, including a nearly complete set of parameters. The numbers shown in the individual cells represent the Spearman correlation coefficients for each combination of parameters presented in the figure. Furthermore, for visual clarity, the color of the cells also illustrates the value of the correlation coefficient, with redder colors indicating stronger positive correlations, whereas bluer colors indicate more negative correlations, according to the shown color bar. The diagonal entries in the matrix all have maximal positive correlations, as these represent the correlation coefficient of a parameter with itself. The empty cells represent combinations of parameters that do not occur in the models. For example, models that include a coronal height $h$ use the \texttt{relxilllp} flavor, which therefore will not contain parameters such as $q_1$, $q_2$, $R_{\rm br}$, $kT_e$, or the disk density $\log N$.

Note that the parameters of the \texttt{zxipcf} and \texttt{apec} components are not included in this matrix, as these spectral components were only included in a handful of spectral fits. For the parameters representing normalizations of components, the ``\_d," ``\_r,'' and ``\_g" indicate that the normalization parameter corresponds to the \texttt{diskbb}, \texttt{relxill}, or \texttt{gaussian} components, while the ``\_A" represents that the shown normalization corresponds to the value determined for the spectrum from the NuSTAR FPMA detector. For visual clarity, the normalizations of the components obtained from fitting the FPMB spectra are not included.

\newpage
\bibliography{paper}{}
\bibliographystyle{aasjournal}

\end{document}